\documentclass[twocolumn,prb,showpacs]{revtex4}
\usepackage{amsmath,amsthm}
\usepackage{epsfig}
\usepackage{graphics}
\usepackage{graphicx}
\usepackage{dcolumn}
\usepackage{multirow}

\usepackage{bm}

\begin{document}

\title{Infrared electrodynamics and ferromagnetism in the topological semiconductors Bi$_2$Te$_3$ and Mn-doped Bi$_2$Te$_3$}

\author{B. C. Chapler,$^{1}$  K. W. Post,$^{1}$  A. R. Richardella,$^{2}$ J. S. Lee,$^{2}$ J. Tao,$^{3}$ N. Samarth,$^{2}$ and D. N. Basov$^{1}$}

\affiliation
{$^{1}$Physics Department, University of California-San Diego, La Jolla, California 92093, USA \\
$^{2}$Department of Physics, The Pennsylvania State University, University Park, Pennsylvania 16802, USA \\
$^{3}$Brookhaven National Laboratory, Upton, New York 11973 USA 
}

\date{\today}

\begin{abstract} 
We report on infrared (IR) optical experiments on Bi$_2$Te$_3$ and Mn-doped Bi$_2$Te$_3$ epitaxial thin films. In the latter film, dilute Mn doping (4.5\%) of the topologically nontrivial semiconductor host results in time-reversal-symmetry-breaking ferromagnetic order below $T_C$=15 K. Our spectroscopic study shows both materials share the Bi$_2$Te$_3$ crystal structure, as well as classification as bulk degenerate semiconductors. Hence the Fermi energy is located in the Bi$_2$Te$_3$ conduction band in both materials, and furthermore, there is no need to invoke topological surface states to describe the conductivity spectra. We also demonstrate that the Drude oscillator strength gives a simple metric with which to distinguish the possibility of topological surface state origins of the low frequency conductance, and conclude that in both the pristine and Mn-doped Bi$_2$Te$_3$ samples the electromagnetic response is indeed dominated by the bulk material properties, rather than those of the surface. An encouraging aspect for taking advantage of the interplay between nontrivial topology and magnetism, however, is that the temperature dependence of the Mn-doped Bi$_2$Te$_3$ film suggests bulk charge carriers do not play a significant role in mediating ferromagnetism. Thus, a truly insulating bulk may still be suitable for the formation of a ferromagnetic ground state in this dilute magnetic topological semiconductor.
\end{abstract}


\maketitle

\section{Introduction}

A topological insulator (TI) is a material with an energy gap in the bulk. However, unlike conventional insulators or semiconductors, TIs are topologically inequivalent to the vacuum. A primary consequence of topological inequivalence is the requirement that the ``gapped region'' in the band structure of the TI is filled with gapless surface state (SS) bands confined to the interface between the bulk TI and a topologically trivial insulator. However, when time reversal symmetry (TRS) is broken, an energy gap can be opened in the SSs. The breaking of TRS is predicted to have a number of interesting consequences in TIs including magnetic monopoles~\cite{Qi2009a}, the topological magnetoelectric effect,~\cite{Qi2008} and quantized Kerr/Faraday rotation~\cite{Tse2010, Maciejko2010}. Research along these lines, for example, has recently led to the prediction~\cite{Yu2010} and first experimental observation of the quantum anomalous Hall effect~\cite{Chang2013}. 

One path to breaking TRS in TIs is to introduce long range ferromagnetic order. Ferromagnetism has been demonstrated in a number of TI candidates doped with transition metal elements~\cite{Dyck2002, Dyck2005, Dyck2003,  Choi2005, Choi2004, Hor2010, Kulbachinskii2002, Zhang2012, Haazen2012, Chang2013a, Checkelsky2012, Xu2012} in this new class of dilute magnetic semiconductors: the dilute magnetic topological insulator. Potential TI phenomena related to the interplay between magnetism and topological SSs is an extremely challenging experimental problem, however. For instance, it remains to be seen if upon transition metal doping, TIs can retain their topological SSs to achieve a gap at the Dirac point. It is theoretically anticipated that this is indeed possible~\cite{Yu2010, Niu2011, Henk2012a, Henk2012}. Experimentally, however, while SSs have been observed in ferromagnetic transition metal-doped TI hosts, they are significantly broadened and not well defined compared to pure samples~\cite{Xu2012, Hor2010}. Furthermore, data have suggested there is a crossover from a TI to a topologically trivial dilute magnetic semiconductor driven by magnetic impurities~\cite{Liu2012}. A further roadblock, and that most relevant to this work, is that TI materials have been plagued by extrinsic defect induced charge carriers. Thus the bulk charge carriers will need to be controlled and eliminated to achieve films that are insulating in the bulk. However, the FM mechanism in canonical dilute mangetic semiconductors such as (Ga,Mn)As and (In,Mn)As is carrier mediated~\cite{MacDonald2005, Sato2010}. A bulk carrier mediated mechanism in transition metal doped TIs would be incompatible with the parallel goal of elimanating bulk charge carriers. 

In our investigation we report the infrared (IR) conductivity and Raman spectra of epitaxial thin films of Bi$_2$Te$_3$ and Mn-doped Bi$_2$Te$_3$ with Curie temperature $T_C$=15 K. IR spectra of a second Mn-doped Bi$_2$Te$_3$ film that does not exhibit ferromagnetism is also discussed (Sec.~\ref{Mnbite}), but is not a focus of this paper. The IR energy range is commensurate with a multitude of electronic processes vital to understanding the physics of these materials. These processes include the excitation of IR active and Raman active phonons, the electrodynamic response of charge carriers, both those in the bulk and potentially those on the surface, optical transitions initiated from mid-gap defect states, and excitations across the bulk energy gap. More importanly, however, IR spectroscopy is a well established tool for investigating ferromagnetism in semiconductors~\cite{Burch2008, Okimoto1995, Chapler2011, Chapler2013, Hirakawa2001}. In particular, carrier mediated mechanisms have specific signatures in the IR.

The inclusion of Bi$_2$Te$_3$ in this work is two-fold. First, the IR response of Bi$_2$Te$_3$ serves as the basis for understanding the effect of magnetic dopants and ferromagnetism on the IR electrodynamics of Mn-doped Bi$_2$Te$_3$, and potentially other related TI candidates. Second, the so called ``second generation materials'' (Bi$_2$Se$_3$, Bi$_2$Te$_3$, and Sb$_2$Te$_3$) are thought to be most promising for unveiling exotic effects predicted for TIs~\cite{Hasan2010}. Improvements in isolating the surface state electrodynamic response in optical experiments has previously been demonstrated in Bi$_2$Se$_3$ thin films, in contrast to bulk crystals~\cite{DiPietro2012, ValdesAguilar2012}. Thus there is clear utility in performing similar IR studies on Bi$_2$Te$_3$ thin films.

Here, spectroscopic features show both Bi$_2$Te$_3$ and Mn-doped Bi$_2$Te$_3$ to be degenerate semiconductors, with no need to invoke topological surface states to describe the IR data. In the case of Bi$_2$Te$_3$, this conclusion is reached despite an optical band gap that is lower than that revealed in photoemission, and a Drude oscillator strength of simlilar magnitude as that observed in Bi$_2$Se$_3$ thin films~\cite{ValdesAguilar2012, Wu2013, Post2013}. Instead, a simple $f$-sum rule based argument is introduced to put an upper limit on the SS conductance and demonstrate that what we observe in Bi$_2$Te$_3$ must be a bulk response. Our experiments indicate a significantly larger bulk charge carrier concentration in the Mn-doped film than that of the pristine Bi$_2$Te$_3$ film. However, despite the large charge carrier concentration, our data suggest bulk charge carriers do not play a significant role in mediating ferromagnetism in Mn-doped Bi$_2$Te$_3$. This latter conclusion is evidenced by the fact that the IR spectrum exhibits remarkably little change upon cooling across the FM transition.

The paper is organized as follows. First, we provide details of our sample growth and initial characterization in Sec.~\ref{samples}. Following that, Sec.~\ref{experimental} describes the experimental methods used in our IR probe. In Sec.~\ref{optics} we present our main results on the IR optical properties of our samples, which is broken in to two sections. Sec.~\ref{bosons} addresses the phonon spectra of our films, while Sec.~\ref{IR} covers the IR electronic response. Discussion of key aspects of these results is found in Sec.~\ref{discussion}. Sec.~\ref{bite} provides conclusions drawn from our data regarding Bi$_2$Te$_3$. Sec.~\ref{Mnbite} considers the effect of magnetic dopants and ferromagnetism on the IR electrodynamics of Mn-doped Bi$_2$Te$_3$. Finally, concluding statements are found in Sec.~\ref{conclusion}.

\section{Samples growth and characterization}
\label{samples}

\begin{figure*}[]
\centering
\includegraphics[width=180mm]{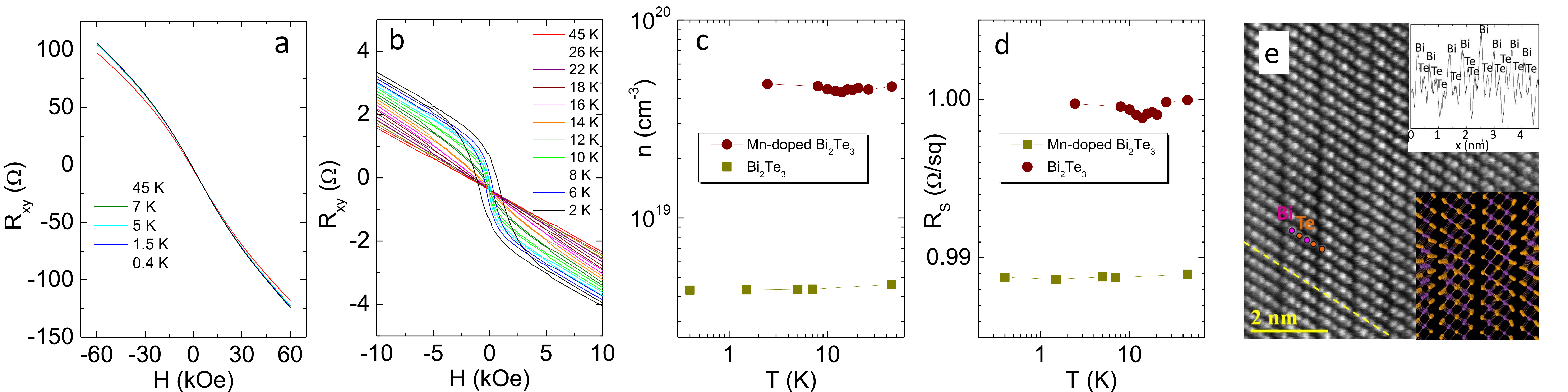}
\caption{(a) Hall resistivity of the Bi$_2$Te$_3$ film. (b) Hall resistivity of the Mn-doped Bi$_2$Te$_3$ film. (c) Electron charge carrier density extracted from the Hall effect for both films. (d) Sheet resistance as a function of temperature for both films. (e) High-angle annular dark-field STEM image of a Mn-doped Bi$_2$Te$_3$ film. Top right inset shows the intensity profile along the dashed yellow line in the main panel. Bottom right inset is a schematic of the Bi$_2$Te$_3$ crystal structure. 
}
\label{TEM}
\end{figure*}

The films in this study were prepared using molecular beam epitaxy (MBE) with the growth direction parallel to the $c$ axis on GaAs (111)B substrates, with film thicknesses of 70 nm and 68 nm for Bi$_2$Te$_3$ and Mn-doped Bi$_2$Te$_3$, respectively. The films are characterized using x-ray diffraction (XRD), Rutherford backscattering (RBS), secondary ion mass spectroscopy (SIMS), scanning transmission electron microscopy (STEM), low temperature magneto-transport and superconducting quantum interference (SQUID) magnetometry. Details of the MBE growth and characterization measurements are provided elsewhere~\cite{Lee2013}. XRD data on the Bi$_2$Te$_3$ film is indicative of typical $c$ axis oriented Bi$_2$Te$_3$. Hall effect data (Fig.~\ref{TEM}a) shows a relatively temperature independent electron charge carrier density $n$ of roughly 4.3$\times$10$^{18}$ cm$^{-3}$ (Fig.~\ref{TEM}c).  

The Mn concentration of the magnetically doped film is 4.5\ atomic \% with 20\% relative error, as determined from RBS and SIMS. The Curie temperature for the onset of ferromagnetism in the Mn-doped Bi$_2$Te$_3$ film is $T_C$=15 K, as established by SQUID magnetometry and by the appearance of a hysteretic anomalous Hall effect below $T_C$ (Fig.~\ref{TEM}b). The Mn-doped Bi$_{2}$Te$_3$ film shows a linear Hall effect above $T_C$, and at magnetic fields outside the hysteretic regime below $T_C$. Fig.~\ref{TEM}c shows the electron charge carrier density extracted from the Hall effect display very little temperature dependence, with a value of roughly 4.5$\times$10$^{19}$ cm$^{-3}$. TEM measurements of the Mn-doped film, completed at Brookhaven National Laboratory using a double Cs corrected microscope, reveal a structure consistent with that of Bi$_2$Te$_3$ (Fig.~\ref{TEM}e). These data also reveal dislocations of the crystal structure to be prevalent in the Mn-doped film. Earlier work on bulk crystals also suggest the possibility of randomly dispersed Bi-bilayers~\cite{Bos2006}. Analysis of the TEM data suggests Mn substitutes either at Bi sites or interstitially, with no obvious evidence of Mn clustering. Though not a main focus this paper, a second Mn-doped Bi$_2$Te$_3$ film that does not exhibit ferromagnetism is discussed in Sec.~\ref{Mnbite}. Based on the growth parameters, and measurements on samples with similar parameters, we estimate that the Mn content is roughly 9\ atomic \% in this latter film. Detailed characterizations of the electronic and crystal structure as a function of Mn concentration can be found in Ref.~\cite{Lee2013}.

\section{Experimental methods}
\label{experimental}

\begin{figure}[]
\centering
\includegraphics[width=86mm]{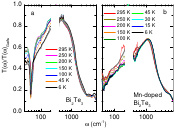}
\caption{The panels show the raw transmission spectra of our samples normalized to that of the GaAs substrate, with Bi$_2$Te$_3$ in panel a and Mn-doped Bi$_2$Te$_3$ in panel b. The thick gray lines are the model fits of the transmission data for extraction of the optical constants, as described in the text, and shown only at room temperature and 6 K for clarity. The data are cut near 300 cm$^{-1}$ due to a phonon in the GaAs substrate that eliminates transmission over this frequency range.
}
\label{rawT}
\end{figure}

The samples were probed optically by normal incidence transmission Fourier transform infrared spectroscopy (FTIR) and Raman spectroscopy ($E$-field $\perp c$ axis). The Raman experiments are performed in the backscattering configuration with a 532 nm laser. In the transmission experiment, unpolarized broad-band IR light with electric field perpendicular to the $c$ axis is incident on the sample. The frequency dependent transmission spectrum is recorded and then normalized by the transmission spectrum of the bare substrate. The raw transmission spectra of both our films, normalized to the GaAs substrate, are shown in Fig.~\ref{rawT}. The transmission spectra are then modeled in order to extract the optical constants of the films, which is imperative to a quantitative understanding of IR data, and is described below.

Transmission spectra are dependent on, aside from thicknesses, both the real and imaginary components of the complex dielectric function $\epsilon(\omega)=\epsilon_1(\omega)+i\epsilon_2(\omega)$. Importantly, these two components are not independent, but linked through the Kramers-Kronig (KK) relations~\cite{Wooten1972}. A convenient method for overcoming the complications of multi-layer systems, and extracting $\epsilon(\omega)$ for a single layer in a multi-layer sample, is via multi-oscillator modeling. In the case of the film on substrate systems studied here, $\epsilon(\omega)$ of the film can be extracted through a KK consistent, multi-oscillator model fit, provided the substrate is measured and modeled separately or $\epsilon(\omega)$ of the substrate is previously determined~\cite{Kuzmenko2005}. By incorporating many oscillators, we make the functional form for the dielectric function more flexible, and thus less model dependent. By constraining the fitting to be KK consistent, we inherit the ability of the KK analysis to extract both the real and imaginary parts of the dielectric function from a single spectrum ($e.g.$ transmission intensity). The fundamental limitations of our technique are thus the experimental error bars and the quality of our least square fitting. This technique has been shown to accurately reproduce optical constants obtained alternately through direct KK analysis of reflectivity, THz time domain spectroscopy, and ellipsometry~\cite{Kuzmenko2005, VanMechelen2008, Jumaily1999}. The IR conductivity spectrum $\sigma(\omega)$ can then be found simply from $\sigma(\omega)=i\frac{\omega(1-\epsilon(\omega))}{60\Omega}$. The thick gray lines in Fig.~\ref{rawT} represent the model fits at room temperature and 6 K. The resulting $\sigma_1(\omega)$ spectra, which describe dissipative processes, are displayed at select temperatures in Fig.~\ref{sigma1}.

\section{optical properties}
\label{optics}
\subsection{Phonon spectrum}
\label{bosons}

From symmetry considerations of the crystal lattice of Bi$_2$Te$_3$ there are 15 lattice dynamical modes at momentum $q$=0: 3 acoustical modes, and 12 optical modes~\cite{Richter1977}. Group theory classifies the 12 optical phonon modes into 2A$_\mathrm{{1g}}$, 2E$_\mathrm{{g}}$, 2A$_\mathrm{{1u}}$, and 2E$_\mathrm{{u}}$. As this system has an inversion center, these optical modes are exclusively Raman- or IR-active, as listed in Table~\ref{phonontable}. The E$_\mathrm{{u}}$ modes are excited by electric fields polarized perpendicular to the $c$-axis, while the A$_\mathrm{{1u}}$ modes are excited only by electric fields polarized parallel to the $c$-axis. Therefore we are not sensitive to the A$_\mathrm{{1u}}$ modes.

\begin{figure}[]
\centering
\includegraphics[width=80mm]{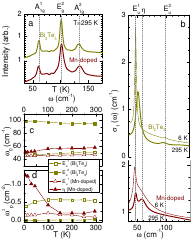}
\caption{(a) Room temperature Raman spectra of our films. Gray dashed lines are guides to the eye to identify categorized Raman active modes (see Table~\ref{phonontable} and discussion in text). (b) Infrared conductivity $\sigma_1(\omega)$ over the phonon region of the spectra of our films. IR data are shown at room temperature and 6 K. Gray dashed lines are again guides to the eye to identify categorized IR active modes (Table~\ref{phonontable} and discussion in text). (c) Temperature dependence of the center frequency ($\omega_{0}$) of IR active phonon modes. (d) Temperature dependence of the oscillator strength ($\omega^2_{p}$) of IR active phonon modes.   
}
\label{phonons}
\end{figure}

The Raman spectra of our films are shown in Fig.~\ref{phonons}a.  Three prominent peaks in each film are clearly observed. The three peaks are readily identified by comparing to the literature as the 2A$_\mathrm{{1g}}$ modes and one E$_\mathrm{{g}}$ mode~\cite{Richter1977}. The peak frequencies of these 3 modes are listed in Table~\ref{phonontable}. We note the other E$_\mathrm{{g}}$ mode, E$^1_\mathrm{{g}}$, is expected near 35 cm$^{-1}$ in Bi$_2$Te$_3$, which is the low frequency limit of our detection. Furthermore, the E$^1_\mathrm{{g}}$ mode, if observed at all in Bi$_2$Te$_3$, has been a very weak feature in the Raman spectrum compared to the 2A$_\mathrm{{1g}}$ modes and the E$^2_\mathrm{{g}}$ mode~\cite{Kullmann1984}. The remarkable similarity of the Raman spectra of the two films supports the conclusion that the Mn-doped film shares the same crystal structure as the Bi$_2$Te$_3$ film.

\begin{table}[]
\caption{Center frequency, in units of cm$^{-1}$, of Raman and IR active phonons observed at room temperature in our Bi$_2$Te$_3$ and Mn-doped Bi$_{2}$Te$_3$ films. These data are compared to those observed in Raman studies of bulk Bi$_2$Te$_3$ crystals in Ref.~\cite{Richter1977}. $^{\dag}$ The value listed for the E$^1_\mathrm{{g}}$ Raman mode is from Ref.~\cite{Kullmann1984}
}
\centering
\begin{tabular}{ c c | c | c | c } 
\hline
& & Bi$_2$Te$_3$ &  Mn-doped & Ref.~\cite{Richter1977} \\ 
\hline\hline 
 Raman & E$^1_\mathrm{g}$ & - & - & 36.5$^{\dag}$  \\
 & A$^1_\mathrm{{1g}}$ & 61 & 60.5 & 62.5  \\
& E$^2_\mathrm{g}$ & 101  & 101.5 & 103  \\
& A$^2_\mathrm{{1g}}$ & 134.5  & 136 & 134   \\
\hline
IR & E$^1_\mathrm{u}$ & 51.31  & 47.2 & 50   \\
&E$^2_\mathrm{u}$ & 94.86  & - & 95   \\
\end{tabular}
\label{phonontable}
\end{table}

The IR active modes detected in our experiments are shown in Fig.~\ref{phonons}b in terms of the IR conductivity ($\sigma_1(\omega)$) at room temperature and 6 K. These data were extracted from our raw transmission data as described in Sec.~\ref{experimental}. The IR active modes modes are all described by Lorentzian oscillators given by,

\begin{equation}
\epsilon(\omega)=\frac{\omega_{p}^2}{\omega_0^2-\omega^2-i\Gamma\omega},
\label{Lorentzian}
\end{equation} 

\noindent where $\omega_p^2$ quantifies the oscillator strength, $\omega_0$ the center frequency, and $\Gamma$ the linewidth. Readily apparent is a distinct mode near 60 cm$^{-1}$ observed in both films, with precise frequencies listed in Table~\ref{phonontable}. This mode is identified as the E$^1_\mathrm{{u}}$ mode. The oscillator strength of the  E$^1_\mathrm{{u}}$ mode in the Bi$_2$Te$_3$ film is much larger than that of the Mn-doped film. Further differences between Bi$_2$Te$_3$ film and the Mn-doped film are noted by the relatively weak E$^1_\mathrm{{u}}$ mode observed near 100 cm$^{-1}$ in Bi$_2$Te$_3$, which is absent in the Mn-doped film (Table~\ref{phonontable}). Moreover, a peak appears in the IR spectra of the Mn-doped film with center frequency $\omega_{0}$ near 60 cm$^{-1}$, which we refer to as the $\eta$ mode. The origin of the $\eta$ mode is unclear. Furthermore, this mode may be a result of a different physical process, such as a low energy interband transition, rather than a phonon. The temperature dependence of $\omega_{0}$ and oscillator strength $\omega^2_{p}$ of the IR active modes observed in our spectra are plotted in Figs.~\ref{phonons}c and d, respectfully. The close agreement of the observed $\omega_{0}$, both in Raman and IR spectra, with those of bulk Bi$_2$Te$_3$ as shown in Table~\ref{phonontable} demonstrates that our epitaxial films share the Bi$_2$Te$_3$ crystal structure.

\subsection{Infrared electrodynamics}
\label{IR}

\begin{figure*}[]
\centering
\includegraphics[width=180mm]{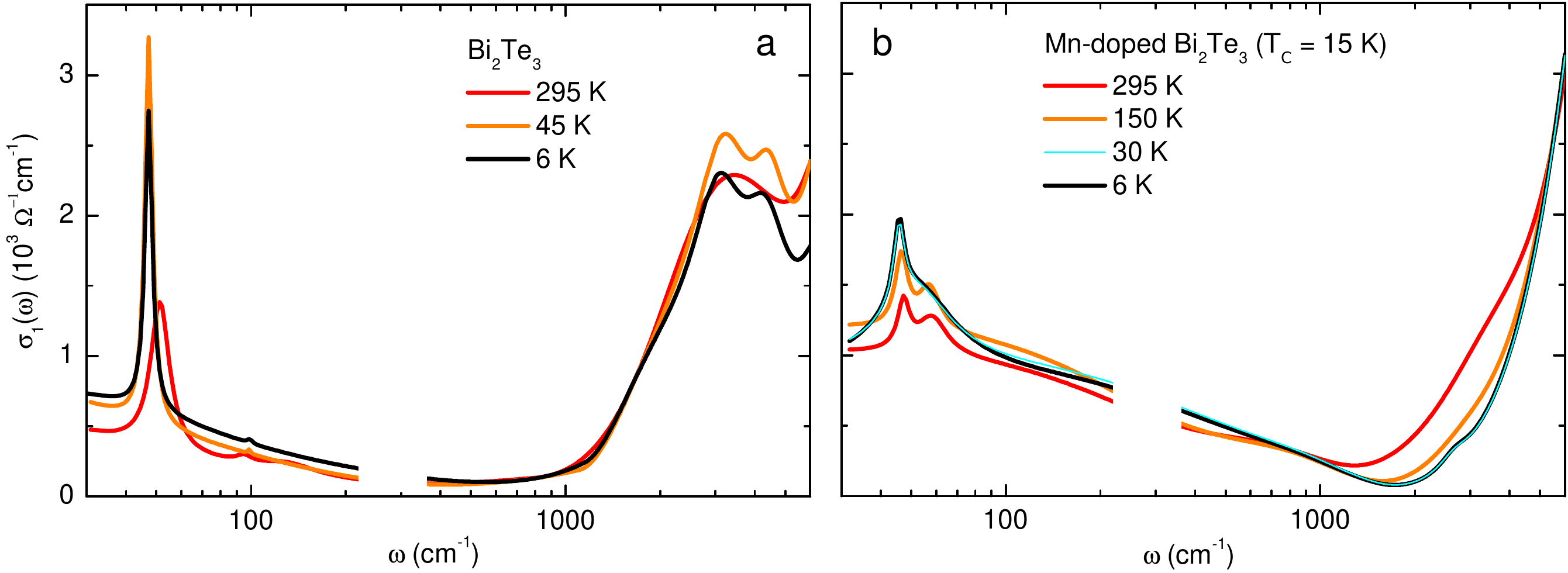}
\caption{Infrared conductivity spectra $\sigma_1(\omega)$ of Bi$_2$Te$_3$ (a), and  the Mn-doped Bi$_2$Te$_3$ sample (b) at select temperatures. The data are cut near 300 cm$^{-1}$ due to a phonon in the GaAs substrate that eliminates transmission over this frequency range.
}
\label{sigma1}
\end{figure*}	

\begin{figure}[]
\centering
\includegraphics[width=80mm]{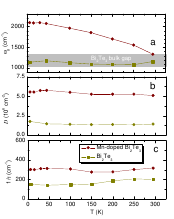}
\caption{Temperature dependence of $\omega_g$ (a), Drude oscillator strength $D$ (b), and free carrier scattering rate 1/$\tau$ (c). The gray region in panel a covers the frequency (energy) range of values reported in the literature for the bulk band gap of Bi$_2$Te$_3$.
}
\label{gapplot}
\end{figure}

The full frequency range of our IR conductivity spectra at select temperatures for the Bi$_2$Te$_3$ and Mn-doped Bi$_2$Te$_3$ films are shown in Figs.~\ref{sigma1}a and b, respectively. We first discuss the spectra of the Bi$_2$Te$_3$ film in Fig.~\ref{sigma1}a. At an energy $\sim$1200 cm$^{-1}$, we observe the onset of optical transitions across the bulk gap of Bi$_2$Te$_3$~\cite{Black1957, Austin1958, Sehr1962, DiPietro2012,LaForge2010, Akrap2012, Dordevic2013, Post2013}, followed by higher energy interband excitations. A broad (relative to phonon modes) Drude-like feature is observed below $\sim$300 cm$^{-1}$ at all temperatures in the Bi$_2$Te$_3$ film. The Drude peak (half-Lorentzian centered at zero frequency) semi-classically describes the characteristic response of free charge carriers in a metal or degenerate semiconductor. As revealed in the figure, the Drude-like feature becomes more prominent upon cooling. The temperature dependence of key parameters of the features described above are shown in Fig.~\ref{gapplot}, and are discussed later.


The $\sigma_1(\omega)$ spectra of the Mn-doped film are shown for select temperatures in Fig.~\ref{sigma1}b. Like the pristine Bi$_2$Te$_3$ film, the Mn-doped sample reveals a clear onset of intergap excitations on the order of 10$^3$ cm$^{-1}$, which sharpens upon cooling. Again similar to the undoped film, there is a Drude-like feature at the intragap frequency scale. Distinct from the pristine Bi$_2$Te$_3$ film, however, is the observation of a broad intragap resonance that lies between the GaAs phonon (where data in the figure is cut) and the onset of intergap excitations in the Mn-doped film.

In Fig.~\ref{gapplot}, we show the temperature dependence of several key parameters of the electrodynamic response of our samples. The method for determining these parameters is discussed sequentially below. We begin by discussing the frequency of the onset of intergap excitations $\omega_g$. The value of $\omega_g$ for our films at all temperatures was determined from linear fits of the square of the imaginary part of the dielectric function $\epsilon_2(\omega)$ near the gap edge. This linear trend of $\epsilon^2_2(\omega)$  well describes the onset of direct interband excitations~\cite{Cardona2010}. In contrast, indirect excitations scale with $\sqrt{\epsilon_2(\omega)}$ near the gap edge~\cite{Cardona2010}. A distinct linear regime in the $\sqrt{\epsilon_2(\omega)}$ spectra was not observed.  

As shown in Fig.~\ref{gapplot}a, $\omega_g$ in the Bi$_2$Te$_3$ film is relatively temperature independent, falling in a range of roughly 1100--1200 cm$^{-1}$. This range of $\omega_g$ is lower that the band gap observed in photoemission experiments~\cite{Chen2009a}, and near the lower end of the range typically extracted for the band gap $E_G$ of Bi$_2$Te$_3$ (1050--1330 cm$^{-1}$ (Refs.~\cite{Harman1957, Sehr1962, Austin1958, Goldsmid1958, Black1957}), represented by the gray bar in Fig.~\ref{gapplot}a). Conversely, Fig.~\ref{gapplot}a shows the Mn-doped film has $\omega_g$ that is larger than $E_G/\hbar$ of Bi$_2$Te$_3$ at all temperatures. The $\omega_g$ feature in the Mn-doped film also exhibits a distinct blue shift as the sample is cooled. At 6K, $\omega_g$ of the Mn-doped film is 765--1045 cm$^{-1}$ larger than $E_G/\hbar$ typically extracted for Bi$_2$Te$_3$, and 950 cm$^{-1}$ larger than the 6 K $\omega_g$ found in our Bi$_2$Te$_3$ film. As will be discussed below, the Mn-doped film also exhibits a significantly larger Drude oscillator strength than that of the pristine Bi$_2$Te$_3$ film. Therefore the correspondingly larger $\omega_g$ values can be understood as a Burstein--Moss shift resulting from $E_F$ residing deeper in the conduction band due to an increase in the carrier density~\cite{Kohler1974}.

We now discuss the temperature dependence of the Drude oscillator strength $D$ of our films, shown in Fig.~\ref{gapplot}b. We quantify $D$ of our films by integration of the intragap spectral weight via the sum rule:

\begin{equation}
D=\frac{30\Omega}{\pi}\int_0^{\omega_c}{\sigma_1(\omega)d\omega},
\label{sumrule}
\end{equation}

\noindent where, guided by the $\omega_g$ values, we use a cutoff frequency $\omega_c$=1000 cm$^{-1}$ for the Bi$_2$Te$_3$ film, and  $\omega_c$=1200 cm$^{-1}$ for the Mn-doped Bi$_2$Te$_3$ film. For the Bi$_2$Te$_3$ film, $D$ stays roughly constant throughout the measured temperature range, with a value of roughly 8$\times$10$^{6}$ cm$^{-2}$. The conservation of Drude oscillator strength with temperature observed in our Bi$_2$Te$_3$ film is typical of metals. This can be understood by realizing that $D$ is directly related to the carrier density $n$, which is largely temperature independent in metallic systems, by $D=\frac{e^2}{4\pi c^2}\frac{n}{m^*}$, where $e$ is the electron charge, $c$ is the speed of light in vacuum, and $m^*$ is the effective carrier mass ($e$, $c$, and $m^*$ are in $cgs$ units).   

Fig.~\ref{gapplot}b shows $D$ of the Mn-doped Bi$_2$Te$_3$ film stays roughly constant throughout the measured temperature range as well. The figure also shows that the Mn-doped film has much larger ($\sim 3\times$) $D$ than that of the Bi$_2$Te$_3$ sample. From the standpoint of $D$ as a measure of $n$, this latter fact is indicative that the Mn-doped film has much larger $n$ than the pristine Bi$_2$Te$_3$ film, consistent with the Hall effect data (Fig.~\ref{TEM}c).

We show the temperature dependence of the free carrier scattering rate 1/$\tau$, in Fig.~\ref{gapplot}c. We quantify $1/\tau$ by

\begin{equation}
1/\tau=\left. \frac{D}{15\Omega}\frac{\sigma_1(\omega)}{\sigma_1(\omega)^2+\sigma_2(\omega)^2} \right|_{\omega'},
\label{scattering}
\end{equation}

\noindent where $\omega'$ is the low frequency cutoff of our data. As can be seen in the figure, 1/$\tau$ of the Bi$_2$Te$_3$ film is roughly 150--200 cm$^{-1}$, and shows a slight narrowing as the film is cooled. This latter trend is characteristic of metallic transport. The Mn-doped sample shows a significantly larger scattering rate ($\sim$300 cm$^{-1}$) than that of the pristine Bi$_2$Te$_3$ film. The enhanced scattering rate with respect to the pristine film is indicative of increased disorder in the Mn-doped film, consistent with the lower mobility extracted from Hall effect measurements.

\section{discussion}
\label{discussion}

\subsection{Bi$_2$Te$_3$}
\label{bite}


It is tempting to consider the possibility that the Drude-like response of the Bi$_2$Te$_3$ sample, or at least a portion of it, originates from topologically protected metallic SSs; a fundamental consequence and principal experimental indicator of nontrivial topology. Such a scenario is even more intriguing given that $\omega_g$ falls within the range of $E_G$ values found in Bi$_2$Te$_3$, suggesting that $E_F$ may actually lie within the bulk gap. Thus it is useful to consider the magnitude of $D$ expected for the Drude response of topological SSs in Bi$_2$Te$_3$. 

In strong TIs, the SSs are predicted~\cite{Li2013} to form Dirac-like bands with a universal background conductance of 0.25$\frac{\pi e^2}{2h}$ (1.52$\times$10$^{-5}$ $\Omega^{-1}$). This background conductance will be completely suppressed, however, at frequencies below 2$E_F$. The $f$-sum rule dictates that this ``missing'' spectral weight from the universal background will appear instead in the Drude portion of the Dirac band conductivity~\cite{Kubo1957, Basov2011, Gusynin2007, Sabio2008}. In this picture, the Drude spectral weight is directly proportional to the location of $E_F$ with respect to the Dirac point. Assuming $E_F$ is 1200 cm$^{-1}$ (0.15 eV) above the Dirac point yields an upper limit for the Drude response of the Dirac band charge carriers, which expressed in 2D units of $D$ is 0.36 cm$^{-1}$. We note that we may observe a Drude response from both top and bottom SSs, suggesting that experimentally a $D$ roughly twice as large ($\sim$0.72 cm$^{-1}$) could still be consistent with the combined Drude response of the Dirac charge carriers from the two surfaces. If the hexagonal warping of the Bi$_2$Te$_3$ SS band structure is taken into account~\cite{Fu2009, Li2013}, this upper limit for $D$ of SSs increases slightly to 0.08 cm$^{-1}$. Averaging $D$ over all measured temperatures in our Bi$_2$Te$_3$ film and expressing the average in 2D units gives 10.2$\pm$1.0 cm$^{-1}$.

From the discussion above, it is clear the observed $D$ is much too large to be considered originating from a topological surface state response, and indicate what we observe is dominated by the bulk. We further note the above argument is idependent of details such as the Fermi velocity, and thus can be applied broadly to TIs. For our data, it could perhaps be considered that the Drude-like response has contributions from both bulk and surface charge carriers. Unfortunately, it is unclear how to uniquely identify or isolate surface state conductance from our measurements. An added complication to the problem of distinguishing topological surface state conduction from that of the bulk is the dramatic band bending in TI materials that has been observed to produce a quantized 2 dimensional electron gas (2DEG) at the surface~\cite{Bianchi2011, King2011}. In any case, it is clear the dominant contribution to the Drude response is from bulk charge carriers, which establishes that $E_F$ resides in the Bi$_2$Te$_3$ conduction band. This latter fact cements the categorization of our Bi$_2$Te$_3$ film as a degenerate semiconductor.

\subsection{Mn-doped Bi$_2$Te$_3$}
\label{Mnbite}

The degenerate semiconductor interpretation is consistent with data for the Mn-doped film as well. This is evidenced by the relatively large $D$ that shows no signs of thermal activation. We also note, the increase in $\omega_g$ with respect to that of pristine Bi$_2$Te$_3$ coupled with the correspondingly larger $D$ in the Mn-doped film implies that there is no reason to expect any discernable contribution to the conduction from topological SSs. A somewhat mysterious aspect of the Mn-doped film is that it is $n$-type rather than $p$-type.  Mn dopants are anticipated both theoretically~\cite{Niu2011} and experimentally~\cite{Hor2010} to substitute for Bi as single acceptors in a Bi$_2$Te$_3$ host. Bi$_2$Te$_3$ films and crystals on the other hand, are typically $n$-type, and this is supported by the Hall and IR data of our Bi$_2$Te$_3$ film.

The IR spectra of Mn-doped Bi$_2$Te$_3$ do show a feature consistent with excitations from mid-gap defect states to unoccupied states above $E_F$ in the conduction band. Namely, there is a broad and relatively weak intragap resonance that lies between the GaAs phonon (where data in the Fig.~\ref{sigma1}b is cut) and the onset of intergap excitations in the Mn-doped film. The width of this broad feature covers an energy scale spanning ranges consistent with optical excitations initiated from both donor $and$ acceptor levels within the bulk band gap. Thus unfortunately, it is difficult to determine if Mn is acting as an acceptor, donor, or is neutral when doped into our films. For instance, the intragap spectral weight represented by the broad resonance could be coming from a combination of Mn acceptor levels and Te vacancy donors and or other unintentional donors.  We speculate that dislocations found in the TEM data of Mn-doped films, which occur at much lower densities in undoped Bi$_2$Te$_3$, may act as an additional source of donor defects resulting in the large electron carrier density of the Mn-doped film.

An additional key observation is that the $\sigma_1(\omega)$ spectra of the Mn-doped Bi$_2$Te$_3$ film exhibits remarkably little change as it is cooled across the FM transition. This is in stark contrast to the canonical dilute magnetic semiconductor Ga$_{1-x}$Mn$_{x}$As. In this latter system, holes originating from Mn acceptors are the principal mediators of ferromagnetism~\cite{MacDonald2005, Sato2010}. Hallmark signs of itinerant FM observed in IR spectra of Ga$_{1-x}$Mn$_{x}$As include an increase of the low energy ``Drude'' spectral weight with the development of magnetization~\cite{Singley2003, Chapler2011}, and scaling of $T_C$ with the IR spectral weight over the intragap energy range~\cite{Singley2002, Chapler2013}. 

\begin{figure}[]
\centering
\includegraphics[width=80mm]{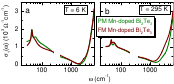}
\caption{Infrared conductivity spectra $\sigma_1(\omega)$ of the FM Mn-doped Bi$_2$Te$_3$ film and a PM Mn-doped Bi$_2$Te$_3$ film at 6 K (a) and room temperature (b). 
}
\label{paramagnetic}
\end{figure}

We do not observe an increase of the low energy spectral weight upon crossing $T_C$ in our Mn-doped sample. This fact is supported by the resistance data in Fig.~\ref{TEM}d. Furthermore, the spectroscopic features of the Mn-doped film are remarkably similar to a second Mn-doped sample we investigated, with the key distinction that the latter sample showed no signs of a ferromagnetic transition down to below 4 K. The $\sigma_1(\omega)$ spectra of the PM Mn-doped film is plotted at room temperature and 6 K  with those of the FM film in Fig.~\ref{paramagnetic}. Key parameters of the electrodynamic response for these two samples at room temperature and 6 K are also shown in Table~\ref{table}. The similarity of the IR response of these two Mn-doped samples, particularly in the intragap region, coupled with the complete absence of ferromagnetism in one of the samples, implies $T_C$ is insensitive to the intragap spectral weight in Mn-doped Bi$_2$Te$_3$. We add that neutron reflectivity measurements are consistent with a uniformly magnetized bulk, and STEM shows no evidence for Mn clustering, supporting that the FM film has true long range FM order~\cite{Lee2013}.  

\begin{table}[]
\caption{Key parameters of the electrodynamic response for the FM and PM Mn-doped Bi$_2$Te$_3$ films, denoted as FM and PM respectively, at 6 K and room temperature.
}
\centering
\begin{tabular}{c | c c | c c } 
\hline
& 6 K & & 295 K & \\ 
\hline\hline 
& PM & FM & PM & FM \\ 
\hline 
$\omega_g$ [cm$^{-1}$]& 2665  & 2095 & 1603 & 1337 \\
$D$ [cm$^{-2}$] & 5.5$\times$10$^6$ & 5.5$\times$10$^6$ & 4.8$\times$10$^6$ & 5.0$\times$10$^6$\\
$1/\tau$ [cm$^{-1}$]& 251 & 303 & 307 & 318 \\
\end{tabular}
\label{table}
\end{table}

A difference between the two Mn-doped samples that may explain the absence of FM in the one is that the PM film was grown with larger Mn content. While the elimination of $T_C$ with an increase in Mn may be initially counter-intuitive, we speculate that the increase of Mn in the PM film may increase the density of dislocations common to Mn-doped Bi$_2$Te$_3$ films, possibly associated to the formation of Bi-bilayers akin to the Bi$_1$Te$_1$ crystal structure~\cite{Bos2006}. This disruption to the translational symmetry and change in electronic properties due to conductive Bi-layers may have reached such an extent as to destroy long range FM order in the PM sample. Detailed characterizations of the electronic and crystal structure as a function of Mn concentration can be found in Ref.~\cite{Lee2013}.

\section{Conclusion and outlook}
\label{conclusion}

Our IR probe of Bi$_2$Te$_3$ and Mn-doped Bi$_2$Te$_3$ epitaxial thin films show both these samples to be degenerate semiconductors. $E_F$ resides just above the conduction band minimum in the Bi$_2$Te$_3$ film, falling within the range of $E_G$ values reported in the literature. The Mn-doped film on the other hand has $E_F$ located roughly 1000 cm$^{-1}$ larger than the Bi$_2$Te$_3$ host conduction band minimum due to a Burstein--Moss shift.

For our Bi$_2$Te$_3$ and Mn-doped Bi$_2$Te$_3$ films, we find no need to invoke topological SSs to describe the IR data. This is likely due to the relatively large bulk charge carrier density that masks or destroys potential hallmarks of SS conduction. Although these signatures are absent from our data, there is evidence that THz/IR probes can be sensitive to topological SS conduction~\cite{Schafgans2012, ValdesAguilar2012, Wu2013, DiPietro2013, Reijnders2014}. Still, other IR experiments have failed to see signatures that could uniquely identify SS conductance~\cite{Dordevic2013, LaForge2010}, even in TI candidates that are quite insulating~\cite{DiPietro2012, Akrap2012, Post2013}. Here we have demonstrated that the Drude oscillator strength gives a simple metric to distinguish the (im)possibility of topological SS origins of the low frequency conductance.
   
Earlier reports have suggested that ferromagnetism in Mn-doped TIs should be analogous to Mn-doped III-V semiconductors~\cite{Hor2010, Checkelsky2012}. However, our results suggest that, unlike in other Mn-doped FM semiconductors~\cite{Burch2008}, charge carriers do not play a significant role in mediating ferromagnetism in Mn-doped Bi$_2$Te$_3$. Alternatively, superexchange~\cite{Niu2011} or an enhanced Van Vleck susceptibility~\cite{Yu2010} have been theoretically proposed as FM mechanisms for TIs doped with transition metal elements. These latter mechanisms do not rely on itinerant charge carriers to mediate FM, and thus can be considered to be consistent with the data from our IR experiments. The bulk charge carriers will of course need to be controlled and eliminated to achieve films that are insulating in the bulk. On this point, however, our data suggests the encouraging possibility that Mn-doped Bi$_2$Te$_3$ may have a FM ground state even when bulk charge carriers are eliminated. The latter proposition is an extremely promising attribute for a dilute magnetic topological semiconductor, as an insulating bulk with broken TRS is required for observation of a number of interesting effects in TIs. 

Sample growth and characterization at Penn State was supported by DARPA (N66001-11-1-4110), ONR (N00014-12-1-0117), ARO-MURI (W911NF-12-1-0461) and the Pennsylvania State University Materials Research Institute Nanofabrication Lab (ECS-0335765). Research at Brookhaven National Laboratory was sponsored by the US Department of Energy (DOE)/Basic Energy Sciences, Materials Sciences and Engineering Division under Contract DE-AC02-98CH10886. Work at UCSD is supported by DARPA.

\maketitle


\begin{thebibliography}{66}
\expandafter\ifx\csname natexlab\endcsname\relax\def\natexlab#1{#1}\fi
\expandafter\ifx\csname bibnamefont\endcsname\relax
  \def\bibnamefont#1{#1}\fi
\expandafter\ifx\csname bibfnamefont\endcsname\relax
  \def\bibfnamefont#1{#1}\fi
\expandafter\ifx\csname citenamefont\endcsname\relax
  \def\citenamefont#1{#1}\fi
\expandafter\ifx\csname url\endcsname\relax
  \def\url#1{\texttt{#1}}\fi
\expandafter\ifx\csname urlprefix\endcsname\relax\def\urlprefix{URL }\fi
\providecommand{\bibinfo}[2]{#2}
\providecommand{\eprint}[2][]{\url{#2}}

\bibitem[{\citenamefont{Qi et~al.}(2009)\citenamefont{Qi, Li, Zang, and
  Zhang}}]{Qi2009a}
\bibinfo{author}{\bibfnamefont{X.-l.} \bibnamefont{Qi}},
  \bibinfo{author}{\bibfnamefont{R.}~\bibnamefont{Li}},
  \bibinfo{author}{\bibfnamefont{J.}~\bibnamefont{Zang}}, \bibnamefont{and}
  \bibinfo{author}{\bibfnamefont{S.-c.} \bibnamefont{Zhang}},
  \textbf{\bibinfo{volume}{323}}, \bibinfo{pages}{1184} (\bibinfo{year}{2009}).

\bibitem[{\citenamefont{Qi et~al.}(2008)\citenamefont{Qi, Hughes, and
  Zhang}}]{Qi2008}
\bibinfo{author}{\bibfnamefont{X.-L.} \bibnamefont{Qi}},
  \bibinfo{author}{\bibfnamefont{T.~L.} \bibnamefont{Hughes}},
  \bibnamefont{and} \bibinfo{author}{\bibfnamefont{S.-C.} \bibnamefont{Zhang}},
  \bibinfo{journal}{Physical Review B} \textbf{\bibinfo{volume}{78}},
  \bibinfo{pages}{195424} (\bibinfo{year}{2008}), ISSN
  \bibinfo{issn}{1098-0121},
  \urlprefix\url{http://link.aps.org/doi/10.1103/PhysRevB.78.195424}.

\bibitem[{\citenamefont{Tse and MacDonald}(2010)}]{Tse2010}
\bibinfo{author}{\bibfnamefont{W.-K.} \bibnamefont{Tse}} \bibnamefont{and}
  \bibinfo{author}{\bibfnamefont{a.~H.} \bibnamefont{MacDonald}},
  \bibinfo{journal}{Physical Review Letters} \textbf{\bibinfo{volume}{105}},
  \bibinfo{pages}{057401} (\bibinfo{year}{2010}), ISSN
  \bibinfo{issn}{0031-9007},
  \urlprefix\url{http://link.aps.org/doi/10.1103/PhysRevLett.105.057401}.

\bibitem[{\citenamefont{Maciejko et~al.}(2010)\citenamefont{Maciejko, Qi, Drew,
  and Zhang}}]{Maciejko2010}
\bibinfo{author}{\bibfnamefont{J.}~\bibnamefont{Maciejko}},
  \bibinfo{author}{\bibfnamefont{X.-L.} \bibnamefont{Qi}},
  \bibinfo{author}{\bibfnamefont{H.~D.} \bibnamefont{Drew}}, \bibnamefont{and}
  \bibinfo{author}{\bibfnamefont{S.-C.} \bibnamefont{Zhang}},
  \bibinfo{journal}{Physical Review Letters} \textbf{\bibinfo{volume}{105}},
  \bibinfo{pages}{166803} (\bibinfo{year}{2010}), ISSN
  \bibinfo{issn}{0031-9007},
  \urlprefix\url{http://link.aps.org/doi/10.1103/PhysRevLett.105.166803}.

\bibitem[{\citenamefont{Yu et~al.}(2010)\citenamefont{Yu, Zhang, Zhang, Zhang,
  Dai, and Fang}}]{Yu2010}
\bibinfo{author}{\bibfnamefont{R.}~\bibnamefont{Yu}},
  \bibinfo{author}{\bibfnamefont{W.}~\bibnamefont{Zhang}},
  \bibinfo{author}{\bibfnamefont{H.-J.} \bibnamefont{Zhang}},
  \bibinfo{author}{\bibfnamefont{S.-C.} \bibnamefont{Zhang}},
  \bibinfo{author}{\bibfnamefont{X.}~\bibnamefont{Dai}}, \bibnamefont{and}
  \bibinfo{author}{\bibfnamefont{Z.}~\bibnamefont{Fang}},
  \bibinfo{journal}{Science (New York, N.Y.)} \textbf{\bibinfo{volume}{329}},
  \bibinfo{pages}{61} (\bibinfo{year}{2010}), ISSN \bibinfo{issn}{1095-9203},
  \urlprefix\url{http://www.ncbi.nlm.nih.gov/pubmed/20522741}.

\bibitem[{\citenamefont{Chang et~al.}(2013{\natexlab{a}})\citenamefont{Chang,
  Zhang, Feng, Shen, Zhang, Guo, Li, Ou, Wei, Wang et~al.}}]{Chang2013}
\bibinfo{author}{\bibfnamefont{C.-Z.} \bibnamefont{Chang}},
  \bibinfo{author}{\bibfnamefont{J.}~\bibnamefont{Zhang}},
  \bibinfo{author}{\bibfnamefont{X.}~\bibnamefont{Feng}},
  \bibinfo{author}{\bibfnamefont{J.}~\bibnamefont{Shen}},
  \bibinfo{author}{\bibfnamefont{Z.}~\bibnamefont{Zhang}},
  \bibinfo{author}{\bibfnamefont{M.}~\bibnamefont{Guo}},
  \bibinfo{author}{\bibfnamefont{K.}~\bibnamefont{Li}},
  \bibinfo{author}{\bibfnamefont{Y.}~\bibnamefont{Ou}},
  \bibinfo{author}{\bibfnamefont{P.}~\bibnamefont{Wei}},
  \bibinfo{author}{\bibfnamefont{L.-L.} \bibnamefont{Wang}},
  \bibnamefont{et~al.}, \bibinfo{journal}{Science (New York, N.Y.)}
  \textbf{\bibinfo{volume}{167}} (\bibinfo{year}{2013}{\natexlab{a}}), ISSN
  \bibinfo{issn}{1095-9203},
  \urlprefix\url{http://www.ncbi.nlm.nih.gov/pubmed/23493424}.

\bibitem[{\citenamefont{Dyck et~al.}(2002)\citenamefont{Dyck, H\'{a}jek,
  Lo\v{s}t'\'{a}k, and Uher}}]{Dyck2002}
\bibinfo{author}{\bibfnamefont{J.}~\bibnamefont{Dyck}},
  \bibinfo{author}{\bibfnamefont{P.}~\bibnamefont{H\'{a}jek}},
  \bibinfo{author}{\bibfnamefont{P.}~\bibnamefont{Lo\v{s}t'\'{a}k}},
  \bibnamefont{and} \bibinfo{author}{\bibfnamefont{C.}~\bibnamefont{Uher}},
  \bibinfo{journal}{Physical Review B} \textbf{\bibinfo{volume}{65}},
  \bibinfo{pages}{115212} (\bibinfo{year}{2002}), ISSN
  \bibinfo{issn}{0163-1829},
  \urlprefix\url{http://link.aps.org/doi/10.1103/PhysRevB.65.115212}.

\bibitem[{\citenamefont{Dyck et~al.}(2005)\citenamefont{Dyck, Dra\v{s}ar,
  Lo\v{s}t'\'{a}k, and Uher}}]{Dyck2005}
\bibinfo{author}{\bibfnamefont{J.}~\bibnamefont{Dyck}},
  \bibinfo{author}{\bibfnamefont{v.}~\bibnamefont{Dra\v{s}ar}},
  \bibinfo{author}{\bibfnamefont{P.}~\bibnamefont{Lo\v{s}t'\'{a}k}},
  \bibnamefont{and} \bibinfo{author}{\bibfnamefont{C.}~\bibnamefont{Uher}},
  \bibinfo{journal}{Physical Review B} \textbf{\bibinfo{volume}{71}},
  \bibinfo{pages}{115214} (\bibinfo{year}{2005}), ISSN
  \bibinfo{issn}{1098-0121},
  \urlprefix\url{http://link.aps.org/doi/10.1103/PhysRevB.71.115214}.

\bibitem[{\citenamefont{Dyck et~al.}(2003)\citenamefont{Dyck, S?vanda,
  Los?t'a´k, Hora´k, Chen, and Uher}}]{Dyck2003}
\bibinfo{author}{\bibfnamefont{J.~S.} \bibnamefont{Dyck}},
  \bibinfo{author}{\bibfnamefont{P.}~\bibnamefont{S?vanda}},
  \bibinfo{author}{\bibfnamefont{P.}~\bibnamefont{Los?t'a´k}},
  \bibinfo{author}{\bibfnamefont{J.}~\bibnamefont{Hora´k}},
  \bibinfo{author}{\bibfnamefont{W.}~\bibnamefont{Chen}}, \bibnamefont{and}
  \bibinfo{author}{\bibfnamefont{C.}~\bibnamefont{Uher}},
  \bibinfo{journal}{Journal of Applied Physics} \textbf{\bibinfo{volume}{94}},
  \bibinfo{pages}{7631} (\bibinfo{year}{2003}), ISSN \bibinfo{issn}{00218979},
  \urlprefix\url{http://link.aip.org/link/JAPIAU/v94/i12/p7631/s1\&Agg=doi}.

\bibitem[{\citenamefont{Choi et~al.}(2005)\citenamefont{Choi, Lee, Kim, Choi,
  Choi, Song, and Cho}}]{Choi2005}
\bibinfo{author}{\bibfnamefont{J.}~\bibnamefont{Choi}},
  \bibinfo{author}{\bibfnamefont{H.-W.} \bibnamefont{Lee}},
  \bibinfo{author}{\bibfnamefont{B.-S.} \bibnamefont{Kim}},
  \bibinfo{author}{\bibfnamefont{S.}~\bibnamefont{Choi}},
  \bibinfo{author}{\bibfnamefont{J.}~\bibnamefont{Choi}},
  \bibinfo{author}{\bibfnamefont{J.~H.} \bibnamefont{Song}}, \bibnamefont{and}
  \bibinfo{author}{\bibfnamefont{S.}~\bibnamefont{Cho}},
  \bibinfo{journal}{Journal of Applied Physics} \textbf{\bibinfo{volume}{97}},
  \bibinfo{pages}{10D324} (\bibinfo{year}{2005}), ISSN
  \bibinfo{issn}{00218979},
  \urlprefix\url{http://link.aip.org/link/JAPIAU/v97/i10/p10D324/s1\&Agg=doi}.

\bibitem[{\citenamefont{Choi et~al.}(2004)\citenamefont{Choi, Choi, Choi, Park,
  Park, Lee, Woo, and Cho}}]{Choi2004}
\bibinfo{author}{\bibfnamefont{J.}~\bibnamefont{Choi}},
  \bibinfo{author}{\bibfnamefont{S.}~\bibnamefont{Choi}},
  \bibinfo{author}{\bibfnamefont{J.}~\bibnamefont{Choi}},
  \bibinfo{author}{\bibfnamefont{Y.}~\bibnamefont{Park}},
  \bibinfo{author}{\bibfnamefont{H.-M.} \bibnamefont{Park}},
  \bibinfo{author}{\bibfnamefont{H.-W.} \bibnamefont{Lee}},
  \bibinfo{author}{\bibfnamefont{B.-C.} \bibnamefont{Woo}}, \bibnamefont{and}
  \bibinfo{author}{\bibfnamefont{S.}~\bibnamefont{Cho}},
  \bibinfo{journal}{Physica Status Solidi (B)} \textbf{\bibinfo{volume}{241}},
  \bibinfo{pages}{1541} (\bibinfo{year}{2004}), ISSN \bibinfo{issn}{03701972},
  \urlprefix\url{http://doi.wiley.com/10.1002/pssb.200304527}.

\bibitem[{\citenamefont{Hor et~al.}(2010)\citenamefont{Hor, Roushan,
  Beidenkopf, Seo, Qu, Checkelsky, Wray, Hsieh, Xia, Xu et~al.}}]{Hor2010}
\bibinfo{author}{\bibfnamefont{Y.~S.} \bibnamefont{Hor}},
  \bibinfo{author}{\bibfnamefont{P.}~\bibnamefont{Roushan}},
  \bibinfo{author}{\bibfnamefont{H.}~\bibnamefont{Beidenkopf}},
  \bibinfo{author}{\bibfnamefont{J.}~\bibnamefont{Seo}},
  \bibinfo{author}{\bibfnamefont{D.}~\bibnamefont{Qu}},
  \bibinfo{author}{\bibfnamefont{J.~G.} \bibnamefont{Checkelsky}},
  \bibinfo{author}{\bibfnamefont{L.~a.} \bibnamefont{Wray}},
  \bibinfo{author}{\bibfnamefont{D.}~\bibnamefont{Hsieh}},
  \bibinfo{author}{\bibfnamefont{Y.}~\bibnamefont{Xia}},
  \bibinfo{author}{\bibfnamefont{S.-Y.} \bibnamefont{Xu}},
  \bibnamefont{et~al.}, \bibinfo{journal}{Physical Review B}
  \textbf{\bibinfo{volume}{81}}, \bibinfo{pages}{195203}
  (\bibinfo{year}{2010}), ISSN \bibinfo{issn}{1098-0121},
  \urlprefix\url{http://link.aps.org/doi/10.1103/PhysRevB.81.195203}.

\bibitem[{\citenamefont{Kulbachinskii et~al.}(2002)\citenamefont{Kulbachinskii,
  Kaminskii, Kindo, Narumi, Suga, Lostak, and Svanda}}]{Kulbachinskii2002}
\bibinfo{author}{\bibfnamefont{V.}~\bibnamefont{Kulbachinskii}},
  \bibinfo{author}{\bibfnamefont{A.}~\bibnamefont{Kaminskii}},
  \bibinfo{author}{\bibfnamefont{K.}~\bibnamefont{Kindo}},
  \bibinfo{author}{\bibfnamefont{Y.}~\bibnamefont{Narumi}},
  \bibinfo{author}{\bibfnamefont{K.}~\bibnamefont{Suga}},
  \bibinfo{author}{\bibfnamefont{P.}~\bibnamefont{Lostak}}, \bibnamefont{and}
  \bibinfo{author}{\bibfnamefont{P.}~\bibnamefont{Svanda}},
  \bibinfo{journal}{Physica B: Condensed Matter}
  \textbf{\bibinfo{volume}{311}}, \bibinfo{pages}{292} (\bibinfo{year}{2002}),
  ISSN \bibinfo{issn}{09214526},
  \urlprefix\url{http://linkinghub.elsevier.com/retrieve/pii/S0921452601009759%
}.

\bibitem[{\citenamefont{Zhang et~al.}(2012)\citenamefont{Zhang, Richardella,
  Rench, Xu, Kandala, Flanagan, Beidenkopf, Yeats, Buckley, Klimov
  et~al.}}]{Zhang2012}
\bibinfo{author}{\bibfnamefont{D.}~\bibnamefont{Zhang}},
  \bibinfo{author}{\bibfnamefont{A.}~\bibnamefont{Richardella}},
  \bibinfo{author}{\bibfnamefont{D.~W.} \bibnamefont{Rench}},
  \bibinfo{author}{\bibfnamefont{S.-Y.} \bibnamefont{Xu}},
  \bibinfo{author}{\bibfnamefont{A.}~\bibnamefont{Kandala}},
  \bibinfo{author}{\bibfnamefont{T.~C.} \bibnamefont{Flanagan}},
  \bibinfo{author}{\bibfnamefont{H.}~\bibnamefont{Beidenkopf}},
  \bibinfo{author}{\bibfnamefont{A.~L.} \bibnamefont{Yeats}},
  \bibinfo{author}{\bibfnamefont{B.~B.} \bibnamefont{Buckley}},
  \bibinfo{author}{\bibfnamefont{P.~V.} \bibnamefont{Klimov}},
  \bibnamefont{et~al.}, \bibinfo{journal}{Physical Review B}
  \textbf{\bibinfo{volume}{86}}, \bibinfo{pages}{205127}
  (\bibinfo{year}{2012}), ISSN \bibinfo{issn}{1098-0121},
  \urlprefix\url{http://link.aps.org/doi/10.1103/PhysRevB.86.205127}.

\bibitem[{\citenamefont{Haazen and Lalo\"{e}}(2012)}]{Haazen2012}
\bibinfo{author}{\bibfnamefont{P.}~\bibnamefont{Haazen}} \bibnamefont{and}
  \bibinfo{author}{\bibfnamefont{J.}~\bibnamefont{Lalo\"{e}}},
  \bibinfo{journal}{Applied Physics Letters} \textbf{\bibinfo{volume}{100}},
  \bibinfo{pages}{082404} (\bibinfo{year}{2012}),
  \urlprefix\url{http://ieeexplore.ieee.org/xpls/abs\_all.jsp?arnumber=6157503%
}.

\bibitem[{\citenamefont{Chang et~al.}(2013{\natexlab{b}})\citenamefont{Chang,
  Zhang, Liu, and Zhang}}]{Chang2013a}
\bibinfo{author}{\bibfnamefont{C.}~\bibnamefont{Chang}},
  \bibinfo{author}{\bibfnamefont{J.}~\bibnamefont{Zhang}},
  \bibinfo{author}{\bibfnamefont{M.}~\bibnamefont{Liu}}, \bibnamefont{and}
  \bibinfo{author}{\bibfnamefont{Z.}~\bibnamefont{Zhang}},
  \bibinfo{journal}{Advanced Materials} pp. \bibinfo{pages}{1065--1070}
  (\bibinfo{year}{2013}{\natexlab{b}}),
  \urlprefix\url{http://onlinelibrary.wiley.com/doi/10.1002/adma.201203493/ful%
l}.

\bibitem[{\citenamefont{Checkelsky et~al.}(2012)\citenamefont{Checkelsky, Ye,
  Onose, Iwasa, and Tokura}}]{Checkelsky2012}
\bibinfo{author}{\bibfnamefont{J.~G.} \bibnamefont{Checkelsky}},
  \bibinfo{author}{\bibfnamefont{J.}~\bibnamefont{Ye}},
  \bibinfo{author}{\bibfnamefont{Y.}~\bibnamefont{Onose}},
  \bibinfo{author}{\bibfnamefont{Y.}~\bibnamefont{Iwasa}}, \bibnamefont{and}
  \bibinfo{author}{\bibfnamefont{Y.}~\bibnamefont{Tokura}},
  \bibinfo{journal}{Nature Physics} \textbf{\bibinfo{volume}{8}},
  \bibinfo{pages}{729} (\bibinfo{year}{2012}), ISSN \bibinfo{issn}{1745-2473},
  \urlprefix\url{http://dx.doi.org/10.1038/nphys2388}.

\bibitem[{\citenamefont{Xu et~al.}(2012)\citenamefont{Xu, Neupane, Liu, Zhang,
  Richardella, Wray, Alidoust, Leandersson, Balasubramanian,
  S\'{a}nchez-barriga et~al.}}]{Xu2012}
\bibinfo{author}{\bibfnamefont{S.-y.} \bibnamefont{Xu}},
  \bibinfo{author}{\bibfnamefont{M.}~\bibnamefont{Neupane}},
  \bibinfo{author}{\bibfnamefont{C.}~\bibnamefont{Liu}},
  \bibinfo{author}{\bibfnamefont{D.}~\bibnamefont{Zhang}},
  \bibinfo{author}{\bibfnamefont{A.}~\bibnamefont{Richardella}},
  \bibinfo{author}{\bibfnamefont{L.~A.} \bibnamefont{Wray}},
  \bibinfo{author}{\bibfnamefont{N.}~\bibnamefont{Alidoust}},
  \bibinfo{author}{\bibfnamefont{M.}~\bibnamefont{Leandersson}},
  \bibinfo{author}{\bibfnamefont{T.}~\bibnamefont{Balasubramanian}},
  \bibinfo{author}{\bibfnamefont{J.}~\bibnamefont{S\'{a}nchez-barriga}},
  \bibnamefont{et~al.}, \bibinfo{journal}{Nature Physics}
  \textbf{\bibinfo{volume}{8}}, \bibinfo{pages}{616} (\bibinfo{year}{2012}),
  ISSN \bibinfo{issn}{1745-2473},
  \urlprefix\url{http://dx.doi.org/10.1038/nphys2351}.

\bibitem[{\citenamefont{Niu et~al.}(2011)\citenamefont{Niu, Dai, Guo, Wei, Ma,
  and Huang}}]{Niu2011}
\bibinfo{author}{\bibfnamefont{C.}~\bibnamefont{Niu}},
  \bibinfo{author}{\bibfnamefont{Y.}~\bibnamefont{Dai}},
  \bibinfo{author}{\bibfnamefont{M.}~\bibnamefont{Guo}},
  \bibinfo{author}{\bibfnamefont{W.}~\bibnamefont{Wei}},
  \bibinfo{author}{\bibfnamefont{Y.}~\bibnamefont{Ma}}, \bibnamefont{and}
  \bibinfo{author}{\bibfnamefont{B.}~\bibnamefont{Huang}},
  \bibinfo{journal}{Applied Physics Letters} \textbf{\bibinfo{volume}{98}},
  \bibinfo{pages}{252502} (\bibinfo{year}{2011}), ISSN
  \bibinfo{issn}{00036951},
  \urlprefix\url{http://link.aip.org/link/APPLAB/v98/i25/p252502/s1\&Agg=doi}.

\bibitem[{\citenamefont{Henk et~al.}(2012{\natexlab{a}})\citenamefont{Henk,
  Flieger, Maznichenko, Mertig, Ernst, Eremeev, and Chulkov}}]{Henk2012a}
\bibinfo{author}{\bibfnamefont{J.}~\bibnamefont{Henk}},
  \bibinfo{author}{\bibfnamefont{M.}~\bibnamefont{Flieger}},
  \bibinfo{author}{\bibfnamefont{I.~V.} \bibnamefont{Maznichenko}},
  \bibinfo{author}{\bibfnamefont{I.}~\bibnamefont{Mertig}},
  \bibinfo{author}{\bibfnamefont{A.}~\bibnamefont{Ernst}},
  \bibinfo{author}{\bibfnamefont{S.~V.} \bibnamefont{Eremeev}},
  \bibnamefont{and} \bibinfo{author}{\bibfnamefont{E.~V.}
  \bibnamefont{Chulkov}}, \bibinfo{journal}{Phys. Rev. Lett.}
  \textbf{\bibinfo{volume}{076801}}, \bibinfo{pages}{3}
  (\bibinfo{year}{2012}{\natexlab{a}}).

\bibitem[{\citenamefont{Henk et~al.}(2012{\natexlab{b}})\citenamefont{Henk,
  Ernst, Eremeev, Chulkov, Maznichenko, and Mertig}}]{Henk2012}
\bibinfo{author}{\bibfnamefont{J.}~\bibnamefont{Henk}},
  \bibinfo{author}{\bibfnamefont{A.}~\bibnamefont{Ernst}},
  \bibinfo{author}{\bibfnamefont{S.}~\bibnamefont{Eremeev}},
  \bibinfo{author}{\bibfnamefont{E.}~\bibnamefont{Chulkov}},
  \bibinfo{author}{\bibfnamefont{I.}~\bibnamefont{Maznichenko}},
  \bibnamefont{and} \bibinfo{author}{\bibfnamefont{I.}~\bibnamefont{Mertig}},
  \bibinfo{journal}{Physical Review Letters} \textbf{\bibinfo{volume}{108}},
  \bibinfo{pages}{206801} (\bibinfo{year}{2012}{\natexlab{b}}), ISSN
  \bibinfo{issn}{0031-9007},
  \urlprefix\url{http://link.aps.org/doi/10.1103/PhysRevLett.108.206801}.

\bibitem[{\citenamefont{Liu et~al.}(2012)\citenamefont{Liu, Zhang, Chang,
  Zhang, Feng, Li, He, Wang, Chen, Dai et~al.}}]{Liu2012}
\bibinfo{author}{\bibfnamefont{M.}~\bibnamefont{Liu}},
  \bibinfo{author}{\bibfnamefont{J.}~\bibnamefont{Zhang}},
  \bibinfo{author}{\bibfnamefont{C.-Z.} \bibnamefont{Chang}},
  \bibinfo{author}{\bibfnamefont{Z.}~\bibnamefont{Zhang}},
  \bibinfo{author}{\bibfnamefont{X.}~\bibnamefont{Feng}},
  \bibinfo{author}{\bibfnamefont{K.}~\bibnamefont{Li}},
  \bibinfo{author}{\bibfnamefont{K.}~\bibnamefont{He}},
  \bibinfo{author}{\bibfnamefont{L.-l.} \bibnamefont{Wang}},
  \bibinfo{author}{\bibfnamefont{X.}~\bibnamefont{Chen}},
  \bibinfo{author}{\bibfnamefont{X.}~\bibnamefont{Dai}}, \bibnamefont{et~al.},
  \bibinfo{journal}{Physical Review Letters} \textbf{\bibinfo{volume}{108}},
  \bibinfo{pages}{036805} (\bibinfo{year}{2012}), ISSN
  \bibinfo{issn}{0031-9007},
  \urlprefix\url{http://link.aps.org/doi/10.1103/PhysRevLett.108.036805}.

\bibitem[{\citenamefont{MacDonald et~al.}(2005)\citenamefont{MacDonald,
  Schiffer, and Samarth}}]{MacDonald2005}
\bibinfo{author}{\bibfnamefont{A.~H.} \bibnamefont{MacDonald}},
  \bibinfo{author}{\bibfnamefont{P.}~\bibnamefont{Schiffer}}, \bibnamefont{and}
  \bibinfo{author}{\bibfnamefont{N.}~\bibnamefont{Samarth}},
  \bibinfo{journal}{Nat. Mater.} \textbf{\bibinfo{volume}{4}},
  \bibinfo{pages}{195} (\bibinfo{year}{2005}),
  \urlprefix\url{http://www.nature.com/nmat/journal/v4/n3/abs/nmat1325.html}.

\bibitem[{\citenamefont{Sato et~al.}(2010)\citenamefont{Sato, Kudrnovsk\'{y},
  Dederichs, Eriksson, Turek, Sanyal, Bouzerar, Katayama-Yoshida, Dinh,
  Fukushima et~al.}}]{Sato2010}
\bibinfo{author}{\bibfnamefont{K.}~\bibnamefont{Sato}},
  \bibinfo{author}{\bibfnamefont{J.}~\bibnamefont{Kudrnovsk\'{y}}},
  \bibinfo{author}{\bibfnamefont{P.~H.} \bibnamefont{Dederichs}},
  \bibinfo{author}{\bibfnamefont{O.}~\bibnamefont{Eriksson}},
  \bibinfo{author}{\bibfnamefont{I.}~\bibnamefont{Turek}},
  \bibinfo{author}{\bibfnamefont{B.}~\bibnamefont{Sanyal}},
  \bibinfo{author}{\bibfnamefont{G.}~\bibnamefont{Bouzerar}},
  \bibinfo{author}{\bibfnamefont{H.}~\bibnamefont{Katayama-Yoshida}},
  \bibinfo{author}{\bibfnamefont{V.~A.} \bibnamefont{Dinh}},
  \bibinfo{author}{\bibfnamefont{T.}~\bibnamefont{Fukushima}},
  \bibnamefont{et~al.}, \bibinfo{journal}{Rev. Mod. Phys.}
  \textbf{\bibinfo{volume}{82}}, \bibinfo{pages}{1633} (\bibinfo{year}{2010}),
  ISSN \bibinfo{issn}{0034-6861},
  \urlprefix\url{http://link.aps.org/doi/10.1103/RevModPhys.82.1633}.

\bibitem[{\citenamefont{Burch et~al.}(2008)\citenamefont{Burch, Awschalom, and
  Basov}}]{Burch2008}
\bibinfo{author}{\bibfnamefont{K.~S.} \bibnamefont{Burch}},
  \bibinfo{author}{\bibfnamefont{D.~D.} \bibnamefont{Awschalom}},
  \bibnamefont{and} \bibinfo{author}{\bibfnamefont{D.~N.} \bibnamefont{Basov}},
  \bibinfo{journal}{J. Magn. Magn. Mater.} \textbf{\bibinfo{volume}{320}},
  \bibinfo{pages}{3207} (\bibinfo{year}{2008}), ISSN \bibinfo{issn}{03048853},
  \urlprefix\url{http://apps.isiknowledge.com/full\_record.do?product=UA\&sear%
ch\_mode=GeneralSearch\&qid=3\&SID=3CBnhmAMGPE1DF4mm17\&page=1\&doc=4\&colname%
=WOS}.

\bibitem[{\citenamefont{Okimoto et~al.}(1995)\citenamefont{Okimoto, Katsufuji,
  Ishikawa, Urushibara, Arima, and Tokura}}]{Okimoto1995}
\bibinfo{author}{\bibfnamefont{Y.}~\bibnamefont{Okimoto}},
  \bibinfo{author}{\bibfnamefont{T.}~\bibnamefont{Katsufuji}},
  \bibinfo{author}{\bibfnamefont{T.}~\bibnamefont{Ishikawa}},
  \bibinfo{author}{\bibfnamefont{A.}~\bibnamefont{Urushibara}},
  \bibinfo{author}{\bibfnamefont{T.}~\bibnamefont{Arima}}, \bibnamefont{and}
  \bibinfo{author}{\bibfnamefont{Y.}~\bibnamefont{Tokura}},
  \bibinfo{journal}{Phys. Rev. Lett.} \textbf{\bibinfo{volume}{75}},
  \bibinfo{pages}{109} (\bibinfo{year}{1995}),
  \urlprefix\url{http://link.aps.org/doi/10.1103/PhysRevLett.75.109}.

\bibitem[{\citenamefont{Chapler et~al.}(2011)\citenamefont{Chapler, Myers,
  Mack, Frenzel, Pursley, Burch, Singley, Dattelbaum, Samarth, Awschalom
  et~al.}}]{Chapler2011}
\bibinfo{author}{\bibfnamefont{B.~C.} \bibnamefont{Chapler}},
  \bibinfo{author}{\bibfnamefont{R.~C.} \bibnamefont{Myers}},
  \bibinfo{author}{\bibfnamefont{S.}~\bibnamefont{Mack}},
  \bibinfo{author}{\bibfnamefont{A.}~\bibnamefont{Frenzel}},
  \bibinfo{author}{\bibfnamefont{B.~C.} \bibnamefont{Pursley}},
  \bibinfo{author}{\bibfnamefont{K.~S.} \bibnamefont{Burch}},
  \bibinfo{author}{\bibfnamefont{E.~J.} \bibnamefont{Singley}},
  \bibinfo{author}{\bibfnamefont{a.~M.} \bibnamefont{Dattelbaum}},
  \bibinfo{author}{\bibfnamefont{N.}~\bibnamefont{Samarth}},
  \bibinfo{author}{\bibfnamefont{D.~D.} \bibnamefont{Awschalom}},
  \bibnamefont{et~al.}, \bibinfo{journal}{Phys. Rev. B}
  \textbf{\bibinfo{volume}{84}}, \bibinfo{pages}{081203}
  (\bibinfo{year}{2011}), ISSN \bibinfo{issn}{1098-0121},
  \urlprefix\url{http://link.aps.org/doi/10.1103/PhysRevB.84.081203}.

\bibitem[{\citenamefont{Chapler et~al.}(2013)\citenamefont{Chapler, Mack,
  Myers, Frenzel, Pursley, Burch, Dattelbaum, Samarth, Awschalom, and
  Basov}}]{Chapler2013}
\bibinfo{author}{\bibfnamefont{B.~C.} \bibnamefont{Chapler}},
  \bibinfo{author}{\bibfnamefont{S.}~\bibnamefont{Mack}},
  \bibinfo{author}{\bibfnamefont{R.~C.} \bibnamefont{Myers}},
  \bibinfo{author}{\bibfnamefont{a.}~\bibnamefont{Frenzel}},
  \bibinfo{author}{\bibfnamefont{B.~C.} \bibnamefont{Pursley}},
  \bibinfo{author}{\bibfnamefont{K.~S.} \bibnamefont{Burch}},
  \bibinfo{author}{\bibfnamefont{a.~M.} \bibnamefont{Dattelbaum}},
  \bibinfo{author}{\bibfnamefont{N.}~\bibnamefont{Samarth}},
  \bibinfo{author}{\bibfnamefont{D.~D.} \bibnamefont{Awschalom}},
  \bibnamefont{and} \bibinfo{author}{\bibfnamefont{D.~N.} \bibnamefont{Basov}},
  \bibinfo{journal}{Physical Review B} \textbf{\bibinfo{volume}{87}},
  \bibinfo{pages}{205314} (\bibinfo{year}{2013}), ISSN
  \bibinfo{issn}{1098-0121},
  \urlprefix\url{http://link.aps.org/doi/10.1103/PhysRevB.87.205314}.

\bibitem[{\citenamefont{Hirakawa}(2001)}]{Hirakawa2001}
\bibinfo{author}{\bibfnamefont{K.}~\bibnamefont{Hirakawa}},
  \bibinfo{journal}{Physica E: Low-dimensional Systems and Nanostructures}
  \textbf{\bibinfo{volume}{10}}, \bibinfo{pages}{215} (\bibinfo{year}{2001}),
  ISSN \bibinfo{issn}{13869477},
  \urlprefix\url{http://linkinghub.elsevier.com/retrieve/pii/S1386947701000856%
}.

\bibitem[{\citenamefont{Hasan and Kane}(2010)}]{Hasan2010}
\bibinfo{author}{\bibfnamefont{M.}~\bibnamefont{Hasan}} \bibnamefont{and}
  \bibinfo{author}{\bibfnamefont{C.}~\bibnamefont{Kane}},
  \bibinfo{journal}{Reviews of Modern Physics} \textbf{\bibinfo{volume}{82}},
  \bibinfo{pages}{3045} (\bibinfo{year}{2010}), ISSN \bibinfo{issn}{0034-6861},
  \urlprefix\url{http://link.aps.org/doi/10.1103/RevModPhys.82.3045}.

\bibitem[{\citenamefont{{Di Pietro} et~al.}(2012)\citenamefont{{Di Pietro},
  Vitucci, Nicoletti, Baldassarre, Calvani, Cava, Hor, Schade, and
  Lupi}}]{DiPietro2012}
\bibinfo{author}{\bibfnamefont{P.}~\bibnamefont{{Di Pietro}}},
  \bibinfo{author}{\bibfnamefont{F.~M.} \bibnamefont{Vitucci}},
  \bibinfo{author}{\bibfnamefont{D.}~\bibnamefont{Nicoletti}},
  \bibinfo{author}{\bibfnamefont{L.}~\bibnamefont{Baldassarre}},
  \bibinfo{author}{\bibfnamefont{P.}~\bibnamefont{Calvani}},
  \bibinfo{author}{\bibfnamefont{R.}~\bibnamefont{Cava}},
  \bibinfo{author}{\bibfnamefont{Y.~S.} \bibnamefont{Hor}},
  \bibinfo{author}{\bibfnamefont{U.}~\bibnamefont{Schade}}, \bibnamefont{and}
  \bibinfo{author}{\bibfnamefont{S.}~\bibnamefont{Lupi}},
  \bibinfo{journal}{Physical Review B} \textbf{\bibinfo{volume}{86}},
  \bibinfo{pages}{045439} (\bibinfo{year}{2012}), ISSN
  \bibinfo{issn}{1098-0121},
  \urlprefix\url{http://link.aps.org/doi/10.1103/PhysRevB.86.045439}.

\bibitem[{\citenamefont{{Vald\'{e}s Aguilar}
  et~al.}(2012)\citenamefont{{Vald\'{e}s Aguilar}, Stier, Liu, Bilbro, George,
  Bansal, Wu, Cerne, Markelz, Oh et~al.}}]{ValdesAguilar2012}
\bibinfo{author}{\bibfnamefont{R.}~\bibnamefont{{Vald\'{e}s Aguilar}}},
  \bibinfo{author}{\bibfnamefont{A.}~\bibnamefont{Stier}},
  \bibinfo{author}{\bibfnamefont{W.}~\bibnamefont{Liu}},
  \bibinfo{author}{\bibfnamefont{L.}~\bibnamefont{Bilbro}},
  \bibinfo{author}{\bibfnamefont{D.}~\bibnamefont{George}},
  \bibinfo{author}{\bibfnamefont{N.}~\bibnamefont{Bansal}},
  \bibinfo{author}{\bibfnamefont{L.}~\bibnamefont{Wu}},
  \bibinfo{author}{\bibfnamefont{J.}~\bibnamefont{Cerne}},
  \bibinfo{author}{\bibfnamefont{A.}~\bibnamefont{Markelz}},
  \bibinfo{author}{\bibfnamefont{S.}~\bibnamefont{Oh}}, \bibnamefont{et~al.},
  \bibinfo{journal}{Physical Review Letters} \textbf{\bibinfo{volume}{108}},
  \bibinfo{pages}{087403} (\bibinfo{year}{2012}), ISSN
  \bibinfo{issn}{0031-9007},
  \urlprefix\url{http://link.aps.org/doi/10.1103/PhysRevLett.108.087403}.

\bibitem[{\citenamefont{Wu et~al.}(2013)\citenamefont{Wu, Brahlek, {Vald\'{e}s
  Aguilar}, Stier, Morris, Lubashevsky, Bilbro, Bansal, Oh, and
  Armitage}}]{Wu2013}
\bibinfo{author}{\bibfnamefont{L.}~\bibnamefont{Wu}},
  \bibinfo{author}{\bibfnamefont{M.}~\bibnamefont{Brahlek}},
  \bibinfo{author}{\bibfnamefont{R.}~\bibnamefont{{Vald\'{e}s Aguilar}}},
  \bibinfo{author}{\bibfnamefont{a.~V.} \bibnamefont{Stier}},
  \bibinfo{author}{\bibfnamefont{C.~M.} \bibnamefont{Morris}},
  \bibinfo{author}{\bibfnamefont{Y.}~\bibnamefont{Lubashevsky}},
  \bibinfo{author}{\bibfnamefont{L.~S.} \bibnamefont{Bilbro}},
  \bibinfo{author}{\bibfnamefont{N.}~\bibnamefont{Bansal}},
  \bibinfo{author}{\bibfnamefont{S.}~\bibnamefont{Oh}}, \bibnamefont{and}
  \bibinfo{author}{\bibfnamefont{N.~P.} \bibnamefont{Armitage}},
  \bibinfo{journal}{Nature Physics} \textbf{\bibinfo{volume}{9}},
  \bibinfo{pages}{410} (\bibinfo{year}{2013}), ISSN \bibinfo{issn}{1745-2473},
  \urlprefix\url{http://www.nature.com/doifinder/10.1038/nphys2647}.

\bibitem[{\citenamefont{Post et~al.}(2013)\citenamefont{Post, Chapler, He, Kou,
  Wang, and Basov}}]{Post2013}
\bibinfo{author}{\bibfnamefont{K.~W.} \bibnamefont{Post}},
  \bibinfo{author}{\bibfnamefont{B.~C.} \bibnamefont{Chapler}},
  \bibinfo{author}{\bibfnamefont{L.}~\bibnamefont{He}},
  \bibinfo{author}{\bibfnamefont{X.}~\bibnamefont{Kou}},
  \bibinfo{author}{\bibfnamefont{K.~L.} \bibnamefont{Wang}}, \bibnamefont{and}
  \bibinfo{author}{\bibfnamefont{D.~N.} \bibnamefont{Basov}},
  \bibinfo{journal}{Phys. Rev. B} \textbf{\bibinfo{volume}{88}},
  \bibinfo{pages}{075121} (\bibinfo{year}{2013}).

\bibitem[{\citenamefont{Lee et~al.}(2013)\citenamefont{Lee, Richardella, Rench,
  and Samarth}}]{Lee2013}
\bibinfo{author}{\bibfnamefont{J.~S.} \bibnamefont{Lee}},
  \bibinfo{author}{\bibfnamefont{A.}~\bibnamefont{Richardella}},
  \bibinfo{author}{\bibfnamefont{D.~W.} \bibnamefont{Rench}}, \bibnamefont{and}
  \bibinfo{author}{\bibfnamefont{N.}~\bibnamefont{Samarth}}, pp.
  \bibinfo{pages}{1--16} (\bibinfo{year}{2013}).

\bibitem[{\citenamefont{Bos et~al.}(2006)\citenamefont{Bos, Lee, Morosan,
  Zandbergen, Lee, Ong, and Cava}}]{Bos2006}
\bibinfo{author}{\bibfnamefont{J.}~\bibnamefont{Bos}},
  \bibinfo{author}{\bibfnamefont{M.}~\bibnamefont{Lee}},
  \bibinfo{author}{\bibfnamefont{E.}~\bibnamefont{Morosan}},
  \bibinfo{author}{\bibfnamefont{H.}~\bibnamefont{Zandbergen}},
  \bibinfo{author}{\bibfnamefont{W.}~\bibnamefont{Lee}},
  \bibinfo{author}{\bibfnamefont{N.}~\bibnamefont{Ong}}, \bibnamefont{and}
  \bibinfo{author}{\bibfnamefont{R.}~\bibnamefont{Cava}},
  \bibinfo{journal}{Physical Review B} \textbf{\bibinfo{volume}{74}},
  \bibinfo{pages}{184429} (\bibinfo{year}{2006}), ISSN
  \bibinfo{issn}{1098-0121},
  \urlprefix\url{http://link.aps.org/doi/10.1103/PhysRevB.74.184429}.

\bibitem[{\citenamefont{Wooten}(1972)}]{Wooten1972}
\bibinfo{author}{\bibfnamefont{F.}~\bibnamefont{Wooten}},
  \emph{\bibinfo{title}{{Optical Properties of Solids}}}
  (\bibinfo{publisher}{Academic}, \bibinfo{address}{New York, London},
  \bibinfo{year}{1972}).

\bibitem[{\citenamefont{Kuzmenko}(2005)}]{Kuzmenko2005}
\bibinfo{author}{\bibfnamefont{A.~B.} \bibnamefont{Kuzmenko}},
  \bibinfo{journal}{Review of Scientific Instruments}
  \textbf{\bibinfo{volume}{76}}, \bibinfo{pages}{083108}
  (\bibinfo{year}{2005}), ISSN \bibinfo{issn}{00346748},
  \urlprefix\url{http://link.aip.org/link/RSINAK/v76/i8/p083108/s1\&Agg=doi}.

\bibitem[{\citenamefont{van Mechelen et~al.}(2008)\citenamefont{van Mechelen,
  van~der Marel, Grimaldi, Kuzmenko, Armitage, Reyren, Hagemann, and
  Mazin}}]{VanMechelen2008}
\bibinfo{author}{\bibfnamefont{J.}~\bibnamefont{van Mechelen}},
  \bibinfo{author}{\bibfnamefont{D.}~\bibnamefont{van~der Marel}},
  \bibinfo{author}{\bibfnamefont{C.}~\bibnamefont{Grimaldi}},
  \bibinfo{author}{\bibfnamefont{a.}~\bibnamefont{Kuzmenko}},
  \bibinfo{author}{\bibfnamefont{N.}~\bibnamefont{Armitage}},
  \bibinfo{author}{\bibfnamefont{N.}~\bibnamefont{Reyren}},
  \bibinfo{author}{\bibfnamefont{H.}~\bibnamefont{Hagemann}}, \bibnamefont{and}
  \bibinfo{author}{\bibfnamefont{I.}~\bibnamefont{Mazin}},
  \bibinfo{journal}{Physical Review Letters} \textbf{\bibinfo{volume}{100}},
  \bibinfo{pages}{226403} (\bibinfo{year}{2008}), ISSN
  \bibinfo{issn}{0031-9007},
  \urlprefix\url{http://link.aps.org/doi/10.1103/PhysRevLett.100.226403}.

\bibitem[{\citenamefont{{J. A. Woolam et al}}(1999)}]{Jumaily1999}
\bibinfo{author}{\bibnamefont{{J. A. Woolam et al}}}, in
  \emph{\bibinfo{booktitle}{Overview of Variable Angle Spectroscopic
  Ellipsometry (VASE), Part I: Basic Theory and Typical APplications}}, edited
  by \bibinfo{editor}{\bibfnamefont{G.~A.} \bibnamefont{Al-Jumaily}}
  (\bibinfo{publisher}{SPIE}, \bibinfo{address}{Denver Colorodo},
  \bibinfo{year}{1999}), p.~\bibinfo{pages}{3},
  \urlprefix\url{http://books.google.com/books?hl=en\&lr=\&id=u15atbXzADUC\&oi%
=fnd\&pg=PR5\&dq=Optical+Metrology\&ots=cc5ZeWpoX\_\&sig=S3tEnSFj4\_ctAJsE0Ktd%
ZqMLASQ}.

\bibitem[{\citenamefont{{Richter, W., Kohler, H., Becker}}(1977)}]{Richter1977}
\bibinfo{author}{\bibfnamefont{C.~R.} \bibnamefont{{Richter, W., Kohler, H.,
  Becker}}}, \bibinfo{journal}{Phys. stat. sol. b}
  \textbf{\bibinfo{volume}{84}}, \bibinfo{pages}{619} (\bibinfo{year}{1977}).

\bibitem[{\citenamefont{Kullmann et~al.}(1984)\citenamefont{Kullmann, Geurts,
  Richter, Rauh, Steigenbbrger, Eichhorn, and Geick}}]{Kullmann1984}
\bibinfo{author}{\bibfnamefont{W.}~\bibnamefont{Kullmann}},
  \bibinfo{author}{\bibfnamefont{J.}~\bibnamefont{Geurts}},
  \bibinfo{author}{\bibfnamefont{W.}~\bibnamefont{Richter}},
  \bibinfo{author}{\bibfnamefont{H.}~\bibnamefont{Rauh}},
  \bibinfo{author}{\bibfnamefont{U.}~\bibnamefont{Steigenbbrger}},
  \bibinfo{author}{\bibfnamefont{G.}~\bibnamefont{Eichhorn}}, \bibnamefont{and}
  \bibinfo{author}{\bibfnamefont{R.}~\bibnamefont{Geick}},
  \bibinfo{journal}{Phys. stat. sol. b} \textbf{\bibinfo{volume}{125}},
  \bibinfo{pages}{131} (\bibinfo{year}{1984}).

\bibitem[{\citenamefont{Black et~al.}(1957)\citenamefont{Black, Conwbll,
  Electric, Seigle, and Spencer}}]{Black1957}
\bibinfo{author}{\bibfnamefont{J.}~\bibnamefont{Black}},
  \bibinfo{author}{\bibfnamefont{E.~M.} \bibnamefont{Conwbll}},
  \bibinfo{author}{\bibfnamefont{S.}~\bibnamefont{Electric}},
  \bibinfo{author}{\bibfnamefont{L.}~\bibnamefont{Seigle}}, \bibnamefont{and}
  \bibinfo{author}{\bibfnamefont{C.~W.} \bibnamefont{Spencer}},
  \bibinfo{journal}{J. Phys. Chem. Solids} \textbf{\bibinfo{volume}{2}},
  \bibinfo{pages}{240} (\bibinfo{year}{1957}).

\bibitem[{\citenamefont{Austin}(1958)}]{Austin1958}
\bibinfo{author}{\bibfnamefont{I.~G.} \bibnamefont{Austin}},
  \bibinfo{journal}{Proc. Phys. Soc.} \textbf{\bibinfo{volume}{72}},
  \bibinfo{pages}{545} (\bibinfo{year}{1958}).

\bibitem[{\citenamefont{{Sehr, R., Testardi}}(1962)}]{Sehr1962}
\bibinfo{author}{\bibfnamefont{L.~R.} \bibnamefont{{Sehr, R., Testardi}}},
  \bibinfo{journal}{J. Phys. Chem. Solids} \textbf{\bibinfo{volume}{23}},
  \bibinfo{pages}{1219} (\bibinfo{year}{1962}).

\bibitem[{\citenamefont{LaForge et~al.}(2010)\citenamefont{LaForge, Frenzel,
  Pursley, Lin, Liu, Shi, and Basov}}]{LaForge2010}
\bibinfo{author}{\bibfnamefont{a.~D.} \bibnamefont{LaForge}},
  \bibinfo{author}{\bibfnamefont{A.}~\bibnamefont{Frenzel}},
  \bibinfo{author}{\bibfnamefont{B.~C.} \bibnamefont{Pursley}},
  \bibinfo{author}{\bibfnamefont{T.}~\bibnamefont{Lin}},
  \bibinfo{author}{\bibfnamefont{X.}~\bibnamefont{Liu}},
  \bibinfo{author}{\bibfnamefont{J.}~\bibnamefont{Shi}}, \bibnamefont{and}
  \bibinfo{author}{\bibfnamefont{D.~N.} \bibnamefont{Basov}},
  \bibinfo{journal}{Physical Review B} \textbf{\bibinfo{volume}{81}},
  \bibinfo{pages}{125120} (\bibinfo{year}{2010}), ISSN
  \bibinfo{issn}{1098-0121},
  \urlprefix\url{http://link.aps.org/doi/10.1103/PhysRevB.81.125120}.

\bibitem[{\citenamefont{Akrap et~al.}(2012)\citenamefont{Akrap, Tran, Ubaldini,
  Teyssier, Giannini, van~der Marel, Lerch, and Homes}}]{Akrap2012}
\bibinfo{author}{\bibfnamefont{A.}~\bibnamefont{Akrap}},
  \bibinfo{author}{\bibfnamefont{M.}~\bibnamefont{Tran}},
  \bibinfo{author}{\bibfnamefont{A.}~\bibnamefont{Ubaldini}},
  \bibinfo{author}{\bibfnamefont{J.}~\bibnamefont{Teyssier}},
  \bibinfo{author}{\bibfnamefont{E.}~\bibnamefont{Giannini}},
  \bibinfo{author}{\bibfnamefont{D.}~\bibnamefont{van~der Marel}},
  \bibinfo{author}{\bibfnamefont{P.}~\bibnamefont{Lerch}}, \bibnamefont{and}
  \bibinfo{author}{\bibfnamefont{C.~C.} \bibnamefont{Homes}},
  \bibinfo{journal}{Physical Review B} \textbf{\bibinfo{volume}{86}},
  \bibinfo{pages}{235207} (\bibinfo{year}{2012}), ISSN
  \bibinfo{issn}{1098-0121},
  \urlprefix\url{http://link.aps.org/doi/10.1103/PhysRevB.86.235207}.

\bibitem[{\citenamefont{Dordevic et~al.}(2013)\citenamefont{Dordevic, Wolf,
  Stojilovic, Lei, and Petrovic}}]{Dordevic2013}
\bibinfo{author}{\bibfnamefont{S.~V.} \bibnamefont{Dordevic}},
  \bibinfo{author}{\bibfnamefont{M.~S.} \bibnamefont{Wolf}},
  \bibinfo{author}{\bibfnamefont{N.}~\bibnamefont{Stojilovic}},
  \bibinfo{author}{\bibfnamefont{H.}~\bibnamefont{Lei}}, \bibnamefont{and}
  \bibinfo{author}{\bibfnamefont{C.}~\bibnamefont{Petrovic}},
  \bibinfo{journal}{Journal of physics. Condensed matter : an Institute of
  Physics journal} \textbf{\bibinfo{volume}{25}}, \bibinfo{pages}{075501}
  (\bibinfo{year}{2013}), ISSN \bibinfo{issn}{1361-648X},
  \urlprefix\url{http://www.ncbi.nlm.nih.gov/pubmed/23328626}.

\bibitem[{\citenamefont{{Yu, Peter Y., Cardona}}(2010)}]{Cardona2010}
\bibinfo{author}{\bibfnamefont{M.}~\bibnamefont{{Yu, Peter Y., Cardona}}},
  \emph{\bibinfo{title}{{Fundamentals of semiconductors}}}
  (\bibinfo{publisher}{Springer-Verlag}, \bibinfo{address}{Berlin},
  \bibinfo{year}{2010}), \bibinfo{edition}{4th} ed., ISBN
  \bibinfo{isbn}{9783642007095}.

\bibitem[{\citenamefont{Chen et~al.}(2009)\citenamefont{Chen, Analytis, Chu,
  Liu, Mo, Qi, Zhang, Lu, Dai, Fang et~al.}}]{Chen2009a}
\bibinfo{author}{\bibfnamefont{Y.~L.} \bibnamefont{Chen}},
  \bibinfo{author}{\bibfnamefont{J.~G.} \bibnamefont{Analytis}},
  \bibinfo{author}{\bibfnamefont{J.-H.} \bibnamefont{Chu}},
  \bibinfo{author}{\bibfnamefont{Z.~K.} \bibnamefont{Liu}},
  \bibinfo{author}{\bibfnamefont{S.-K.} \bibnamefont{Mo}},
  \bibinfo{author}{\bibfnamefont{X.~L.} \bibnamefont{Qi}},
  \bibinfo{author}{\bibfnamefont{H.~J.} \bibnamefont{Zhang}},
  \bibinfo{author}{\bibfnamefont{D.~H.} \bibnamefont{Lu}},
  \bibinfo{author}{\bibfnamefont{X.}~\bibnamefont{Dai}},
  \bibinfo{author}{\bibfnamefont{Z.}~\bibnamefont{Fang}}, \bibnamefont{et~al.},
  \bibinfo{journal}{Science (New York, N.Y.)} \textbf{\bibinfo{volume}{325}},
  \bibinfo{pages}{178} (\bibinfo{year}{2009}), ISSN \bibinfo{issn}{1095-9203},
  \urlprefix\url{http://www.ncbi.nlm.nih.gov/pubmed/19520912}.

\bibitem[{\citenamefont{Harman et~al.}(1957)\citenamefont{Harman, Paris,
  Miller, and Goefung}}]{Harman1957}
\bibinfo{author}{\bibfnamefont{T.~C.} \bibnamefont{Harman}},
  \bibinfo{author}{\bibfnamefont{B.}~\bibnamefont{Paris}},
  \bibinfo{author}{\bibfnamefont{S.~E.} \bibnamefont{Miller}},
  \bibnamefont{and} \bibinfo{author}{\bibfnamefont{H.~L.}
  \bibnamefont{Goefung}}, \bibinfo{journal}{J. Phys. Chem. Solids}
  \textbf{\bibinfo{volume}{2}}, \bibinfo{pages}{181} (\bibinfo{year}{1957}).

\bibitem[{\citenamefont{Goldsmid}(1958)}]{Goldsmid1958}
\bibinfo{author}{\bibfnamefont{H.~J.} \bibnamefont{Goldsmid}},
  \bibinfo{journal}{Proc. Phys. Soc.} \textbf{\bibinfo{volume}{71}},
  \bibinfo{pages}{633} (\bibinfo{year}{1958}).

\bibitem[{\citenamefont{Kohler and Hartmann}(1974)}]{Kohler1974}
\bibinfo{author}{\bibfnamefont{H.}~\bibnamefont{Kohler}} \bibnamefont{and}
  \bibinfo{author}{\bibfnamefont{J.}~\bibnamefont{Hartmann}},
  \bibinfo{journal}{Phys. Stat. sol. (b)} \textbf{\bibinfo{volume}{63}},
  \bibinfo{pages}{171} (\bibinfo{year}{1974}).

\bibitem[{\citenamefont{Li and Carbotte}(2013)}]{Li2013}
\bibinfo{author}{\bibfnamefont{Z.}~\bibnamefont{Li}} \bibnamefont{and}
  \bibinfo{author}{\bibfnamefont{J.~P.} \bibnamefont{Carbotte}},
  \bibinfo{journal}{Physical Review B} \textbf{\bibinfo{volume}{87}},
  \bibinfo{pages}{155416} (\bibinfo{year}{2013}), ISSN
  \bibinfo{issn}{1098-0121},
  \urlprefix\url{http://link.aps.org/doi/10.1103/PhysRevB.87.155416}.

\bibitem[{\citenamefont{Kubo}(1957)}]{Kubo1957}
\bibinfo{author}{\bibfnamefont{R.}~\bibnamefont{Kubo}},
  \bibinfo{journal}{Journal of the Physical Society of Japan}
  \textbf{\bibinfo{volume}{12}}, \bibinfo{pages}{570} (\bibinfo{year}{1957}).

\bibitem[{\citenamefont{Basov et~al.}(2011)\citenamefont{Basov, Averitt,
  van~der Marel, Dressel, and Haule}}]{Basov2011}
\bibinfo{author}{\bibfnamefont{D.}~\bibnamefont{Basov}},
  \bibinfo{author}{\bibfnamefont{R.}~\bibnamefont{Averitt}},
  \bibinfo{author}{\bibfnamefont{D.}~\bibnamefont{van~der Marel}},
  \bibinfo{author}{\bibfnamefont{M.}~\bibnamefont{Dressel}}, \bibnamefont{and}
  \bibinfo{author}{\bibfnamefont{K.}~\bibnamefont{Haule}},
  \bibinfo{journal}{Rev. Mod. Phys.} \textbf{\bibinfo{volume}{83}},
  \bibinfo{pages}{471} (\bibinfo{year}{2011}), ISSN \bibinfo{issn}{0034-6861},
  \urlprefix\url{http://link.aps.org/doi/10.1103/RevModPhys.83.471}.

\bibitem[{\citenamefont{Gusynin et~al.}(2007)\citenamefont{Gusynin, Sharapov,
  and Carbotte}}]{Gusynin2007}
\bibinfo{author}{\bibfnamefont{V.}~\bibnamefont{Gusynin}},
  \bibinfo{author}{\bibfnamefont{S.}~\bibnamefont{Sharapov}}, \bibnamefont{and}
  \bibinfo{author}{\bibfnamefont{J.}~\bibnamefont{Carbotte}},
  \bibinfo{journal}{Physical Review B} \textbf{\bibinfo{volume}{75}},
  \bibinfo{pages}{165407} (\bibinfo{year}{2007}), ISSN
  \bibinfo{issn}{1098-0121},
  \urlprefix\url{http://link.aps.org/doi/10.1103/PhysRevB.75.165407}.

\bibitem[{\citenamefont{Sabio et~al.}(2008)\citenamefont{Sabio, Nilsson, and
  {Castro Neto}}}]{Sabio2008}
\bibinfo{author}{\bibfnamefont{J.}~\bibnamefont{Sabio}},
  \bibinfo{author}{\bibfnamefont{J.}~\bibnamefont{Nilsson}}, \bibnamefont{and}
  \bibinfo{author}{\bibfnamefont{a.}~\bibnamefont{{Castro Neto}}},
  \bibinfo{journal}{Physical Review B} \textbf{\bibinfo{volume}{78}},
  \bibinfo{pages}{075410} (\bibinfo{year}{2008}), ISSN
  \bibinfo{issn}{1098-0121},
  \urlprefix\url{http://link.aps.org/doi/10.1103/PhysRevB.78.075410}.

\bibitem[{\citenamefont{Fu}(2009)}]{Fu2009}
\bibinfo{author}{\bibfnamefont{L.}~\bibnamefont{Fu}},
  \bibinfo{journal}{Physical Review Letters} \textbf{\bibinfo{volume}{103}},
  \bibinfo{pages}{266801} (\bibinfo{year}{2009}), ISSN
  \bibinfo{issn}{0031-9007},
  \urlprefix\url{http://link.aps.org/doi/10.1103/PhysRevLett.103.266801}.

\bibitem[{\citenamefont{Bianchi et~al.}(2011)\citenamefont{Bianchi, Hatch, Mi,
  Iversen, and Hofmann}}]{Bianchi2011}
\bibinfo{author}{\bibfnamefont{M.}~\bibnamefont{Bianchi}},
  \bibinfo{author}{\bibfnamefont{R.~C.} \bibnamefont{Hatch}},
  \bibinfo{author}{\bibfnamefont{J.}~\bibnamefont{Mi}},
  \bibinfo{author}{\bibfnamefont{B.~B.} \bibnamefont{Iversen}},
  \bibnamefont{and} \bibinfo{author}{\bibfnamefont{P.}~\bibnamefont{Hofmann}},
  \bibinfo{journal}{Physical Review Letters} \textbf{\bibinfo{volume}{107}},
  \bibinfo{pages}{086802} (\bibinfo{year}{2011}), ISSN
  \bibinfo{issn}{0031-9007},
  \urlprefix\url{http://link.aps.org/doi/10.1103/PhysRevLett.107.086802}.

\bibitem[{\citenamefont{King et~al.}(2011)\citenamefont{King, Hatch, Bianchi,
  Ovsyannikov, Lupulescu, Landolt, Slomski, Dil, Guan, Mi et~al.}}]{King2011}
\bibinfo{author}{\bibfnamefont{P.~D.~C.} \bibnamefont{King}},
  \bibinfo{author}{\bibfnamefont{R.~C.} \bibnamefont{Hatch}},
  \bibinfo{author}{\bibfnamefont{M.}~\bibnamefont{Bianchi}},
  \bibinfo{author}{\bibfnamefont{R.}~\bibnamefont{Ovsyannikov}},
  \bibinfo{author}{\bibfnamefont{C.}~\bibnamefont{Lupulescu}},
  \bibinfo{author}{\bibfnamefont{G.}~\bibnamefont{Landolt}},
  \bibinfo{author}{\bibfnamefont{B.}~\bibnamefont{Slomski}},
  \bibinfo{author}{\bibfnamefont{J.~H.} \bibnamefont{Dil}},
  \bibinfo{author}{\bibfnamefont{D.}~\bibnamefont{Guan}},
  \bibinfo{author}{\bibfnamefont{J.~L.} \bibnamefont{Mi}},
  \bibnamefont{et~al.}, \bibinfo{journal}{Physical Review Letters}
  \textbf{\bibinfo{volume}{107}}, \bibinfo{pages}{096802}
  (\bibinfo{year}{2011}), ISSN \bibinfo{issn}{0031-9007},
  \urlprefix\url{http://link.aps.org/doi/10.1103/PhysRevLett.107.096802}.

\bibitem[{\citenamefont{Singley et~al.}(2003)\citenamefont{Singley, Burch,
  Kawakami, Stephens, Awschalom, and Basov}}]{Singley2003}
\bibinfo{author}{\bibfnamefont{E.~J.} \bibnamefont{Singley}},
  \bibinfo{author}{\bibfnamefont{K.~S.} \bibnamefont{Burch}},
  \bibinfo{author}{\bibfnamefont{R.~K.} \bibnamefont{Kawakami}},
  \bibinfo{author}{\bibfnamefont{J.}~\bibnamefont{Stephens}},
  \bibinfo{author}{\bibfnamefont{D.~D.} \bibnamefont{Awschalom}},
  \bibnamefont{and} \bibinfo{author}{\bibfnamefont{D.~N.} \bibnamefont{Basov}},
  \bibinfo{journal}{Phys. Rev. B} \textbf{\bibinfo{volume}{68}},
  \bibinfo{pages}{165204} (\bibinfo{year}{2003}), ISSN
  \bibinfo{issn}{0163-1829},
  \urlprefix\url{http://link.aps.org/doi/10.1103/PhysRevB.68.165204}.

\bibitem[{\citenamefont{Singley et~al.}(2002)\citenamefont{Singley, Kawakami,
  Awschalom, and Basov}}]{Singley2002}
\bibinfo{author}{\bibfnamefont{E.~J.} \bibnamefont{Singley}},
  \bibinfo{author}{\bibfnamefont{R.~K.} \bibnamefont{Kawakami}},
  \bibinfo{author}{\bibfnamefont{D.~D.} \bibnamefont{Awschalom}},
  \bibnamefont{and} \bibinfo{author}{\bibfnamefont{D.~N.} \bibnamefont{Basov}},
  \bibinfo{journal}{Phys. Rev. Lett.} \textbf{\bibinfo{volume}{89}},
  \bibinfo{pages}{097203} (\bibinfo{year}{2002}), ISSN
  \bibinfo{issn}{0031-9007},
  \urlprefix\url{http://link.aps.org/doi/10.1103/PhysRevLett.89.097203}.

\bibitem[{\citenamefont{Schafgans et~al.}(2012)\citenamefont{Schafgans, Post,
  Taskin, Ando, Qi, Chapler, and Basov}}]{Schafgans2012}
\bibinfo{author}{\bibfnamefont{A.~A.} \bibnamefont{Schafgans}},
  \bibinfo{author}{\bibfnamefont{K.}~\bibnamefont{Post}},
  \bibinfo{author}{\bibfnamefont{A.}~\bibnamefont{Taskin}},
  \bibinfo{author}{\bibfnamefont{Y.}~\bibnamefont{Ando}},
  \bibinfo{author}{\bibfnamefont{X.-L.} \bibnamefont{Qi}},
  \bibinfo{author}{\bibfnamefont{B.}~\bibnamefont{Chapler}}, \bibnamefont{and}
  \bibinfo{author}{\bibfnamefont{D.}~\bibnamefont{Basov}},
  \bibinfo{journal}{Physical Review B} \textbf{\bibinfo{volume}{85}},
  \bibinfo{pages}{3} (\bibinfo{year}{2012}), ISSN \bibinfo{issn}{1098-0121},
  \urlprefix\url{http://link.aps.org/doi/10.1103/PhysRevB.85.195440}.

\bibitem[{\citenamefont{{Di Pietro} et~al.}(2013)\citenamefont{{Di Pietro},
  Ortolani, Limaj, {Di Gaspare}, Giliberti, Giorgianni, Brahlek, Bansal,
  Koirala, Oh et~al.}}]{DiPietro2013}
\bibinfo{author}{\bibfnamefont{P.}~\bibnamefont{{Di Pietro}}},
  \bibinfo{author}{\bibfnamefont{M.}~\bibnamefont{Ortolani}},
  \bibinfo{author}{\bibfnamefont{O.}~\bibnamefont{Limaj}},
  \bibinfo{author}{\bibfnamefont{a.}~\bibnamefont{{Di Gaspare}}},
  \bibinfo{author}{\bibfnamefont{V.}~\bibnamefont{Giliberti}},
  \bibinfo{author}{\bibfnamefont{F.}~\bibnamefont{Giorgianni}},
  \bibinfo{author}{\bibfnamefont{M.}~\bibnamefont{Brahlek}},
  \bibinfo{author}{\bibfnamefont{N.}~\bibnamefont{Bansal}},
  \bibinfo{author}{\bibfnamefont{N.}~\bibnamefont{Koirala}},
  \bibinfo{author}{\bibfnamefont{S.}~\bibnamefont{Oh}}, \bibnamefont{et~al.},
  \bibinfo{journal}{Nature nanotechnology} \textbf{\bibinfo{volume}{8}},
  \bibinfo{pages}{556} (\bibinfo{year}{2013}), ISSN \bibinfo{issn}{1748-3395},
  \urlprefix\url{http://www.ncbi.nlm.nih.gov/pubmed/23872838}.

\bibitem[{\citenamefont{Reijnders et~al.}(2014)\citenamefont{Reijnders, Tian,
  Sandilands, Pohl, Kivlichan, Zhao, Jia, Charles, Cava, Alidoust
  et~al.}}]{Reijnders2014}
\bibinfo{author}{\bibfnamefont{A.~a.} \bibnamefont{Reijnders}},
  \bibinfo{author}{\bibfnamefont{Y.}~\bibnamefont{Tian}},
  \bibinfo{author}{\bibfnamefont{L.~J.} \bibnamefont{Sandilands}},
  \bibinfo{author}{\bibfnamefont{G.}~\bibnamefont{Pohl}},
  \bibinfo{author}{\bibfnamefont{I.~D.} \bibnamefont{Kivlichan}},
  \bibinfo{author}{\bibfnamefont{S.~Y.~F.} \bibnamefont{Zhao}},
  \bibinfo{author}{\bibfnamefont{S.}~\bibnamefont{Jia}},
  \bibinfo{author}{\bibfnamefont{M.~E.} \bibnamefont{Charles}},
  \bibinfo{author}{\bibfnamefont{R.~J.} \bibnamefont{Cava}},
  \bibinfo{author}{\bibfnamefont{N.}~\bibnamefont{Alidoust}},
  \bibnamefont{et~al.}, \bibinfo{journal}{Physical Review B}
  \textbf{\bibinfo{volume}{89}}, \bibinfo{pages}{075138}
  (\bibinfo{year}{2014}), ISSN \bibinfo{issn}{1098-0121},
  \urlprefix\url{http://link.aps.org/doi/10.1103/PhysRevB.89.075138}.

\end{thebibliography}
\end{document}